\documentclass{aastex62}

\usepackage{graphicx}	
\usepackage{amsmath}	
\usepackage{amssymb}	
\usepackage[mathscr]{euscript}
\usepackage{float}

\def\kgm{{\rm kg~m$^{-3}$}}
\newcommand{\gps}{\ensuremath{g_{\rm P1}}}
\newcommand{\rps}{\ensuremath{r_{\rm P1}}}
\newcommand{\ips}{\ensuremath{i_{\rm P1}}}
\newcommand{\zps}{\ensuremath{z_{\rm P1}}}
\newcommand{\yps}{\ensuremath{y_{\rm P1}}}
\newcommand{\wps}{\ensuremath{w_{\rm P1}}}
\newcommand{\grizy}{g_{\rm P1}\rps\ips\zps\yps}

\newcommand{\PS}{\protect \hbox {Pan-STARRS}}

\graphicspath{{./}{figures/}}

\received{January 1, 2018}
\revised{January 7, 2018}
\accepted{\today}
\submitjournal{ApJ}

\shorttitle{Extreme Asteroids in Pan-STARRS 1}
\shortauthors{McNeill et al.}

\begin{document}

\title{Extreme asteroids in the Pan-STARRS 1 Survey}

\correspondingauthor{Andrew McNeill}
\email{andrew.mcneill@nau.edu}

\author{Andrew McNeill}
\affil{Department of Physics and Astronomy, Northern Arizona University, Flagstaff, AZ 86011, USA}
\affil{Astrophysics Research Centre, Queen's University Belfast, BT7 1NN, Northern Ireland, UK}

\author{Alan Fitzsimmons}
\affiliation{Astrophysics Research Centre, Queen's University Belfast, BT7 1NN, Northern Ireland, UK}

\author{Robert Jedicke}
\affiliation{Institute for Astronomy, University of Hawaii at Manoa, Honolulu, HI 96822, USA}

\author{Pedro Lacerda}
\affiliation{Astrophysics Research Centre, Queen's University Belfast, BT7 1NN, Northern Ireland, UK}

\author{Eva Lilly}
\affiliation{Institute for Astronomy, University of Hawaii at Manoa, Honolulu, HI 96822, USA}

\author{Andrew Thompson}
\affiliation{Astrophysics Research Centre, Queen's University Belfast, BT7 1NN, Northern Ireland, UK}

\author{David E. Trilling}
\affiliation{Department of Physics and Astronomy, Northern Arizona University, Flagstaff, AZ 86011, USA}

\author{Ernst DeMooij}
\affiliation{School of Physical Sciences, Dublin City University, Glasnevin, Dublin 9, Ireland}

\author{Matthew J. Hooton}
\affiliation{Astrophysics Research Centre, Queen's University Belfast, BT7 1NN, Northern Ireland, UK}

\author{Christopher A. Watson}
\affiliation{Astrophysics Research Centre, Queen's University Belfast, BT7 1NN, Northern Ireland, UK}



\begin{abstract}

Using the first 18 months of the Pan-STARRS 1 survey we have identified 33 candidate high-amplitude objects for follow-up observations and carried out observations of 22 asteroids. 4 of the observed objects were found to have observed amplitude $A_{obs}\geq 1.0$ mag. We find that these high amplitude objects are most simply explained by single rubble pile objects with some density-dependent internal strength, allowing them to resist mass shedding even at their highly elongated shapes. 3 further objects although below the cut-off for 'high-amplitude' had a combination of elongation and rotation period which also may require internal cohesive strength, depending on the density of the body. We find that none of the 'high-amplitude asteroids' identified here require any unusual cohesive strengths to resist rotational fission. 3 asteroids were sufficiently observed to allow for shape and spin pole models to be determined through light curve inversion. 45864 was determined to have retrograde rotation with spin pole axes $\lambda=218\pm 10^{\circ}, \beta=-82\pm 5^{\circ}$ and asteroid 206167 was found to have best fit spin pole axes $\lambda= 57 \pm 5^{\circ}$, $\beta=-67 \pm 5^{\circ}$. An additional object not initially measured with $A_{obs}>1.0$ mag, 49257, was determined to have a shape model which does suggest a high-amplitude object. Its spin pole axes were best fit for values $\lambda=112\pm 6^{\circ}, \beta=6\pm 5^{\circ}$. In the course of this project to date no large super-fast rotators ($P_{rot} < 2.2$ h) have been identified.

\end{abstract}



\section{Introduction} \label{sec:intro}

We classify 'extreme' asteroids based on their shape and spin state data. In this research we define an 'extreme asteroid' as an object rotating with a period shorter than the spin barrier ($P<2.2$ h), or an object with measured light curve amplitude $A\geq 1.0$ mag. This cut-off for 'high-amplitude' is largely arbitrary as to date there has been no work calculating a limit for rubble pile asteroids accounting for angle of friction.

\begin{figure}
  \begin{center}
\includegraphics[width=0.5\textwidth]{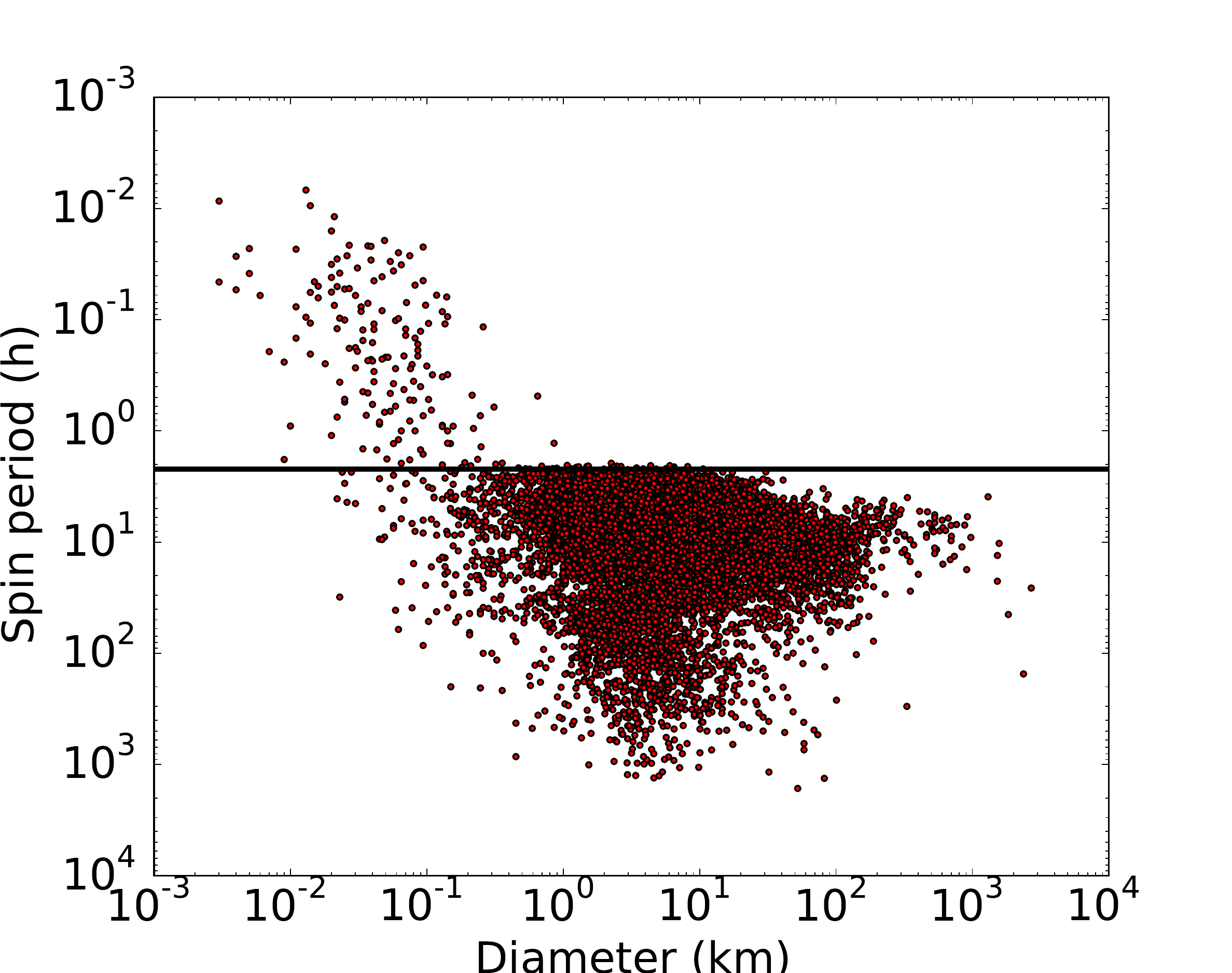}
   \caption{A plot of spin period against diameter for all objects in the Light Curve Database with quality code $\geq 2-$. The black line represents the 'spin barrier' at $P=2.2$ h corresponding to the critical spin rate for an object with bulk density $\rho=2500$\kgm. Data from (\citealt{warner2009}; last updated 6 September 2016)}
\label{fig:lcdb_spin}
  \end{center}
 \end{figure}

Figure~\ref{fig:lcdb_spin} shows the rotation rate of all main belt asteroids plotted against diameter as recorded in the Light Curve Database (\citealt{warner2009}; last updated 6 September 2016). In this plot we use only objects which have been assigned a quality code $U\geq 2-$. The quality code is assigned according to the reliability of the period result obtained. A value of $U=1$ is assigned for a period determined from a light curve fragment which can not be considered reliable. $U=2$ corresponds to a result from partial light curves with a period value known within an uncertainty of $30\%$. $U=3$ is assigned to accurate results from full light curve coverage (\citealt{warner2009}).

In Figure~\ref{fig:lcdb_spin} there is a clear cut-off in rotation rate at period $P\approx 2.2$ h. This is approximately the critical spin rate for a strengthless body assuming a bulk density $\rho=2500$ \kgm. The critical spin rate is defined as the speed of rotation required for the body to undergo rotational disruption. This suggests that most asteroids in the $0.2<D<10$ km size range are loosely bound, 'rubble piles' of aggregate material with zero tensile strength.

\subsection{Large Super-Fast Rotators}

Large super fast rotators (SFRs) are defined as asteroids with diameter $D>200$ m and with spin periods shorter than the spin barrier at $P=2.2$ h. This critical spin period is an estimate based on an average density value for asteroids and as such is not a hard cut-off, likewise the diameter cut-off should be considered empirical. Literature values for the spin barrier range from $2\leq P_{bar} < 2.2$ h (\citealt{pravec2002}) and at the time of writing there were 20 objects in the LCDB not considered as SFRs but with rotation periods in this range. A higher than average density will allow for faster rotation without disruption and without requiring the object itself to be especially unusual. For this reason we only consider objects with periods clearly shorter than $P=2.2$ h.

The first large SFR was discovered in 2001 by \cite{pravec2002}. Asteroid 2001OE84 was found to have a rotation period $P=29.1909\pm 0.0001$ min. The spectra obtained for this object suggested it to be a stony S-type object. Using the average albedo for this type of asteroid its diameter is estimated at $D=0.7$ km. Since the discovery of 2001OE84 there have been three further confirmed SFRs in the main asteroid belt along with a series of further candidates unconfirmed at the time this was written. Two of these have been identified by the Palomar Transient Factory (\citealt{chang2015}; \citealt{chang2016}). The confirmed SFRs are 2005UW163 and 1999RE88. 2005UW163 is an object in the inner main asteroid belt with an estimated mean diameter $D=0.6\pm 0.3$ km and a measured rotation period $P=1.29\pm 0.06$ h. 1999RE88 is also found in the inner main belt with diameter $D=1.9\pm 0.3$ km and spin period $P=1.96\pm 0.01$ h. \cite{polishook2016} present the final main belt SFR, 2000GD65 which is an S-type inner main belt object with $D=2.3\pm 0.6$km and $P=1.9529\pm 0.0002$ h making it the largest SFR known to date. A further Near Earth Asteroid (NEA) has also been found to have a superfast rotation period. \cite{rozitis2014} show that 1950DA has diameter $D=1.30\pm 0.13$ km and rotation period $P=2.12$ h, only slightly below the critical spin rate stated here of $P=2.2$ h.

If an object was monolithic then this would imply significant cohesive strength and would allow for faster rotation than the spin barrier. It is unlikely that any monoliths of this size remain intact in the asteroid belt without having undergone further collisional disruption, however, \cite{polishook2017} have argued that a monolithic asteroid best explains the rotational behaviour of 2001OE84 due to the high cohesive strengths which would be otherwise required to explain its short rotation period.

\subsection{Highly elongated objects}
\label{sub:elong}

Assuming the simplest case of a strengthless triaxial ellipsoid there is a limit on the elongation the object can have before it becomes unstable and will undergo rotational fission (\citealt{jeans1919}). For a triaxial ellipsoid ($a\geq b\geq c$) this limit is at $\frac{b}{a}=0.43$. Assuming a constant albedo for the asteroid and that all brightness variation is solely due to the projected shape of the body about its rotation axis this corresponds to a light curve amplitude of $A\leq 0.9$ mag. However, there are several asteroids in the size range $0.2<D<10$ km where objects are predominantly rubble piles showing light curve amplitudes larger than this. Objects showing amplitudes significantly greater than this therefore can not simply be assumed to be strengthless ellipsoids. As in the case of SFRs, a cohesive strength within the objects may have to be accounted for. Assuming rubble pile objects to have some cohesive strength between their components may also allow for greater elongation. 

It is not expected that asteroids in this size range can be approximated entirely as fluidic bodies. The use of hydrostatic equilibrium models in this case is not used to obtain accurate shape estimates for the objects, instead this is a useful approximation which allows us then to account for the presence of cohesive strength in a body of this approximate shape and size.

A difference in the shape of the object may also account for this change. Although triaxial ellipsoids become unstable where $\frac{b}{a}<0.43$ there is a sequence of equilibrium figures in the form of bilobed objects which allow for a lower $\frac{b}{a}$ while maintaining object stability (\citealt{chandrasekhar1969}; \cite{descamps2015}). A bilobed object is described by the four axes $a\geq b\geq c\geq c'$ where $c'$ is the minimum length of the c-axis at a 'waist' between two larger lobes. The shape can best be described as like that of a dumb-bell or dog-bone.

We consider an object in the form of a Jacobi triaxial ellipsoid ($a>b \ge c$). As an object of this shape rotates around its axis the projected area seen by an observer will change. This produces a double peaked light curve when measured. The variation in the light curve can be used as a means of estimating the axis ratios of the asteroid, as shown in Equation~\ref{eqn:deltam} where $A_{obs}$ represents the observed amplitude of the light curve, $\theta$ is the latitude of the spin pole axis. Assuming the spin axis of the object to be perpendicular to the orbital plane i.e. $\theta=90^{\circ}$, this reduces to Equation~\ref{eqn:deltam2}.

\begin{equation} 
A_{obs}=-2.5log\frac{b}{a}-1.25log{\frac{a^{2}cos^{2}\theta + c^{2}sin^{2}\theta}{b^{2}cos^{2}\theta+c^{2}sin^{2}\theta}}
\label{eqn:deltam}
\end{equation}

\begin{equation} 
A_{obs}=-2.5log\frac{b}{a}
\label{eqn:deltam2}
\end{equation}
 
The shapes of Jacobi ellipsoids can be determined by setting $a=1$ and solving Equation~\ref{eqn:chandraparty} from \cite{chandrasekhar1969} for relative axis ratios $b$ and $c$ where $\Delta$ is given in Equation~\ref{eqn:Delta}.

\begin{equation} 
a^{2}b^{2}\int_{0}^{\infty}\frac{du}{(a^{2}+u)(b^{2}+u)\Delta}=c^{2}\int_{0}^{\infty}\frac{du}{(c^{2}+u)\Delta}
\label{eqn:chandraparty}
\end{equation}

\begin{equation} 
\Delta^{2}=(a^{2}+u)(b^{2}+u)(c^{2}+u)
\label{eqn:Delta}
\end{equation}

If these $a$, $b$ and $c$ values are known along with the angular rotation frequency of the object, $\omega$ and assuming the object to have no internal strength then the density of the asteroid may be estimated using Equation~\ref{eqn:density2} where G is the gravitational constant and $\rho$ is the density of the body in kg m$^{-3}$, assuming that the density of the object is constant throughout.

\begin{equation} 
\frac{\omega^{2}}{\pi G\rho}=2abc\int_{0}^{\infty}\frac{u du}{(a^{2}+u)(b^{2}+u)\Delta}
\label{eqn:density2}
\end{equation}

Figures producing $A_{obs} > 0.9$ mag would be unstable and will undergo mass loss during rotation unless some significant internal strength is assumed (\citealt{jeans1919}). Therefore we assume that it is unlikely that our highest amplitude targets can be simply explained by an ellipsoid of this nature, however, some objects closer to the $1.0$ mag selection criterion may be explained by Jacobi ellipsoids possessing cohesive strength and thus we can consider this a valid shape for at least some of our targets. 

However, to assume these rubble pile objects to be simply fluid ellipsoids would be incorrect. Instead we also account for the effect of the angle of sliding friction between the constituent parts of the rubble pile. To constrain the potential cohesive strengths of these rotating ellipsoids we use a simplified form of the Drucker-Prager model, a stress-strain model commonly used in the study of geological materials (\citealt{alejano2012}). The Drucker-Prager failure criterion is a three-dimensional model estimating the stresses within a geological material at its critical rotation state. The shear stresses on a body in three orthogonal $xyz$ axes are dependent on the shape, density and rotational properties of the body (\citealt{holsapple2007}).

\begin{equation} 
\sigma_{x}=(\rho\omega^{2}-2\pi\rho^{2}GA_{x})\frac{a^{2}}{5}
\label{eqn:sigmax}
\end{equation}

\begin{equation} 
\sigma_{y}=(\rho\omega^{2}-2\pi\rho^{2}GA_{y})\frac{b^{2}}{5}
\label{eqn:sigmay}
\end{equation}

\begin{equation} 
\sigma_{z}=(-2\pi\rho^{2}GA_{z})\frac{c^{2}}{5}
\label{eqn:sigmaz}
\end{equation}

The three $A_{i}$ functions are dimensionless parameters dependent on the axis ratios of the body.

\begin{equation} 
A_{x}=\frac{c}{a}\frac{b}{a}\int_{0}^{\infty}\frac{1}{(u+1)^{3/2}(u+\frac{b}{a}^{2})^{1/2}(u+\frac{c}{a}^{2})^{1/2}}du
\label{eqn:Ax}
\end{equation}

\begin{equation} 
A_{y}=\frac{c}{a}\frac{b}{a}\int_{0}^{\infty}\frac{1}{(u+1)^{1/2}(u+\frac{b}{a}^{2})^{3/2}(u+\frac{c}{a}^{2})^{1/2}}du
\label{eqn:Ay}
\end{equation}

\begin{equation} 
A_{z}=\frac{c}{a}\frac{b}{a}\int_{0}^{\infty}\frac{1}{(u+1)^{1/2}(u+\frac{b}{a}^{2})^{1/2}(u+\frac{c}{a}^{2})^{3/2}}du
\label{eqn:Az}
\end{equation}

The Drucker-Prager failure criterion defines the point at which the object will break up and is given in Equation~\ref{eqn:drucker} where $k$ is the internal cohesive strength of the body and $s$ is a slope parameter dependent on the assumed angle of friction, $\phi$. The expression for s is given in Equation~\ref{eqn:slope}.

\begin{equation} 
\frac{1}{6}[(\sigma_{x}-\sigma_{y})^{2}+(\sigma_{y}-\sigma_{z})^{2}+(\sigma_{z}-\sigma_{x})^{2}] \leq [k-s(\sigma_{x}+\sigma_{y}+\sigma_{z})]^{2}
\label{eqn:drucker}
\end{equation}

\begin{equation} 
s=\frac{2\rm{sin}\phi}{\sqrt{3}(3-\rm{sin}\phi)}
\label{eqn:slope}
\end{equation}

Using a simple model based on this failure criterion we determine the required cohesive strength as a function of density for each of the objects we have observed using input parameters determined from the observations. For the cases where high density objects require cohesive strength this is not necessarily to resist rotational fission, rather that some strength is required for the object to not undergo reshaping toward a more spherical shape. We use a Monte Carlo numerical simulation to determine solutions for a range of values using the uncertainties in each of the asteroid parameters to constrain the required cohesion of each object. Potential phase angle opposition effects for different asteroid taxonomies are also accounted for in these uncertainties.

\cite{holsapple2001} show that in their model most asteroids can be explained with an angle of friction of $15^{\circ}$ or less. We assume this value in our calculations of density, it should be noted that these density values are simply estimates of the required density assuming this angle of sliding friction and should not be taken to be directly determined values.

\subsection{Contact binary or bilobed objects}

Contact binary or bilobed objects have been discovered in almost every small body population in the solar system. The abundance of binary systems can be relatively high with $10-20\%$ of Near Earth Object (NEO) and Kuiper Belt Object (KBO) populations showing evidence of binarity (\citealt{pravec2006}; \citealt{lacerda2011}). A subset of the binary population exist as contact binary or bilobed systems. 3 bilobed objects have been identified in the Trojan family (\citealt{mann2007}) and 2 in the Kuiper Belt (\citealt{sheppard2004}; \citealt{lacerda2014}). Contact binary or bilobed systems are important as by modelling their photometric light curves it is possible to obtain estimates of their bulk densities, assuming a rubble-pile structure. Spectroscopic, colour and albedo data allow the composition of an asteroid to be determined. With values of grain densities for different mineralogies known from meteorite studies, the bulk porosity of the object can be calculated. Also, non-contact binaries will be more readily collisionally disrupted than contact binaries. Hence, the ratio of bilobed bodies and non-contact binaries can be used to constrain the collisional statistics and history of asteroids in the main belt (\citealt{nesvorny2011}).

The formation mechanism of binary asteroids is dependent on the size of the system. Large binary objects ($D>40$ km) are thought to be formed in collisional events while small binaries ($D<20$ km) may also be explained by rotational fission due to the YORP effect (\citealt{bottke2002}; \citealt{pravec2010}). \cite{benner2006} define an asteroid contact binary as an asteroid consisting of two lobes in contact with a bimodal mass distribution that may once have been separate. Contact binary asteroids form whenever a separated binary spirals inward and the components collide "gently" due to either tidal interactions between the bodies or some angular momentum drain. This will occur when a separated binary has no possible stable synchronous orbit (\citealt{taylor2011}).



Hektor is the most notable bilobed object in the Solar System. It is a D-type Jupiter Trojan with $D=233$ km (\citealt{cruikshank2001}; \citealt{usui2011}). \cite{dunlap1969} obtained light curve data for this object and found it to have amplitude $A_{obs}> 1.0$ mag. \cite{cook1971} first proposed the idea that this object was best explained as a close or contact binary system rather than simply as an elongated object, a conclusion later backed up by \cite{weidenschilling1980}. Recent studies have shown that $>10\%$ of Trojans may be contact binaries, a similar proportion to the Kuiper Belt and Near Earth Object populations (\citealt{pravec2006}; \citealt{lacerda2011}). However, no such abundance has been observed in the main asteroid belt. To date 6 potential contact binary or bilobed objects have been found in the main belt (\citealt{ostro2000}; \citealt{marchis2005}; \citealt{descamps2007}; \citealt{shepard2015}). Only one of these objects falls in the size range primarily dealt with in this paper ($D<10$ km). (3169) Ostro is a $5.15\pm 0.09$ km object orbiting with a semi major axis $a=1.89$AU and with a rotation period $P=6.509\pm 0.001$h (\citealt{descamps2007}; \citealt{mainzer2011}). The light curve obtained for this object by \cite{descamps2007} showed amplitude $A_{obs}=1.20\pm 0.05$ mag. Modelling the object as a Roche binary they find a solution fitting both their own observations and previous apparitions for an object with a bulk density $\rho=2600$ \kgm. At present Ostro is the best candidate for a small main belt bilobed object.

Binary objects can produce much larger light curve variations than would be possible for a single object (\citealt{leone1984}). In this case the light curve variation will be produced as the binary system orbits its barycentre and the primary and secondary objects occult one another. As no eclipses are visible in the light curves of the high amplitude asteroids observed, this suggests that if these objects are indeed binaries then they must be close binaries or potentially contact binaries. We first consider a simplified case of two spheres in contact. 









The maximum light curve amplitude a bilobed object of this form can produce is $A=0.75$ mag, well below the cut off we set for 'high-amplitude asteroids' ($A=1.0$ mag). However, this is a highly simplified model as if we consider the components of the binary to be strengthless 'rubble-piles' then gravitational forces between the two will cause deformation of the bodies allowing for even greater light curve amplitudes. To attempt to account for this deformation we use the Roche binary approximation (\citealt{chandra1963}; \citealt{leone1984}).

\subsection{Roche ellipsoid constraints}

Roche's problem deals with the effect of tidal forces acting between two bodies rotating around a common centre of mass in a binary system. In this problem the secondary component of the binary is treated as a rigid sphere and the primary's deformation to an equilibrium shape by gravitational forces determined, the resulting shape defined as a Roche ellipsoid, and then vice versa. Using the Roche binary approximation we assume a binary system consisting of these two Roche ellipsoids. The resulting shape of each component can be calculated assuming two tidally locked components with the same rotation rate. In the case of close components with mass ratio $p\approx 1$ this approximation will introduce the greatest uncertainty into the result. In this case the elongation of the primary and secondary will be underestimated, causing an underestimation of the light curve amplitude.

From \cite{chandra1963} the axis ratios of a Roche ellipsoid component with a given value of $p$ can be determined from Equation~\ref{eqn:roche} where $a,b$ and $c$ are the lengths of the three axes of the ellipsoid and $p$ is the mass ratio. The parameters $A_{i}$ are defined by \cite{chandra1962} and are given in Equations~\ref{eqn:a1}, \ref{eqn:a2} and \ref{eqn:a3}.

\begin{equation} 
\frac{(3+p)a^{2}+c^{2}}{pb^{2}+c^{2}}=\frac{A_{1}a^{2}-A_{3}c^{2}}{A_{2}b^{2}-A_{3}c^{2}}
\label{eqn:roche}
\end{equation}

\begin{equation} 
A_{1}=\frac{2}{a^{3}\rm{sin}^{3}\phi}\frac{1}{\rm{sin}^{2}\theta}[F(\theta, \phi) - E(\theta, \phi)]
\label{eqn:a1}
\end{equation}

\begin{equation} 
A_{2}=\frac{2}{a^{3}\rm{sin}^{3}\phi}\frac{1}{\rm{sin}^{2}\rm{cos}^{2}\theta}[E(\theta, \phi) - F(\theta, \phi){\rm{cos}}^{2}\theta - \frac{c}{b}\rm{sin}^{2}\theta\rm{sin}\phi]
\label{eqn:a2}
\end{equation}

\begin{equation} 
A_{3}=\frac{2}{a^{3}\rm{sin}^{3}\phi}\frac{1}{\rm{cos}^{2}\theta}[\frac{b}{c}\rm{sin}\phi - E(\theta, \phi)]
\label{eqn:a3}
\end{equation}

$E(\theta, \phi)$ and $F(\theta, \phi)$ are the standard incomplete elliptical integrals of the first and second kind given in Equations~\ref{eqn:etp} and \ref{eqn:ftp} with arguments $\theta$ and $\phi$ as defined in Equations~\ref{eqn:theta} and \ref{eqn:phi}. 

\begin{equation} 
E(\theta, \phi) = \int_{0}^{\phi}(1-\rm{sin}^{2}\theta\rm{sin}^{2}\phi)^{\frac{1}{2}} d\phi
\label{eqn:etp}
\end{equation}

\begin{equation} 
F(\theta, \phi) = \int_{0}^{\phi}(1-\rm{sin}^{2}\theta\rm{sin}^{2}\phi)^{-\frac{1}{2}} d\phi
\label{eqn:ftp}
\end{equation}

\begin{equation} 
\theta=\rm{sin}^{-1}(\sqrt{\frac{a^{2}-b^{2}}{a^{2}-c^{2}}})
\label{eqn:theta}
\end{equation}

\begin{equation} 
\phi= {\rm{cos}}^{-1} (\frac{c}{a})
\label{eqn:phi}
\end{equation}

The model of \cite{lacerda2007} sets three of the four parameters $p$, $a$, $b$ and $c$ and interpolates to solve for the fourth. For the primary component the major axis length is set to $a=1$ and the axis length $b$ is found for each combination of $p$ and $c$ between a minimum value and $1.00$ in iterative steps of $0.01$ (\citealt{lacerda2007} use a mass ratio value $q$ corresponding to $1/p$). This process is repeated using reciprocal values for the mass ratio to determine the shape of the secondary. For the axis ratios determined using this method the square of the rotational frequency $\omega$ is given in units of $\pi G\rho$ by Equation~\ref{eqn:rochedens} where $\rho$ is the bulk density of the ellipsoid.

\begin{equation} 
\frac{1}{1+p}\frac{\omega^{2}}{\pi G\rho}=2abc\frac{A_{1}a^{2}-A_{3}c^{2}}{(3+p)a^{2}+c^{2}}
\label{eqn:rochedens}
\end{equation}

\cite{lacerda2007} defined a valid Roche binary solution as when the ($p,a,b,c$) values for both the primary and secondary components give the same result when substituted into the right hand side of Equation~\ref{eqn:rochedens}.  A Roche binary system can produce a lightcurve amplitude of $1.5$ mag, a factor of two larger than for a binary composed of spherical bodies.

It is worth noting that to conclude any binarity or bilobed nature from a single set of light curve observations as described here is speculative. To determine the true shape of these objects further observations would be required. Solutions calculated from Equation 22 should be considered viable but should not be mistaken for a definite statement on the binarity of these objects.

\section{Pan-STARRS Data}

The \PS 1 1.8m telescope on Haleakala contains a 1.4 gigapixel orthogonal transfer array CCD camera (GPC1) and covers a field of view of $\sim 7$ square degrees on the sky (\citealt{denneau2013}; \citealt{chambers2016}). GPC1 is made up of an $8$x$8$ grid of OTA CCDs with each of these OTAs in turn comprising of an $8$x$8$ array of $590$ x $598$ $10 \mu$m CCDs (\citealt{tonry2012}). The exposure time of observations is survey/filter dependent and the CCD readout time is approximately $7$s  (\citealt{Magnier2013}; \citealt{magnier2016}).

The $\grizy$$\wps$ filter system used by the survey is similar to the Sloan-Gunn system, however, there are slight differences in wavelength range (\citealt{tonry2012}). In this work we deal exclusively with data taken using the $\wps$-band filter. The $\wps$ filter spans the combined range of the $\gps$, $\rps$ and $\ips$ filters allowing detections down to 22nd magnitude.

The initial target list was constructed using the first 18 months of PanSTARRS-1 survey data. We consider objects in the range $2<a<4$ AU with an upper limit of $e<0.35$ to prevent any contamination of the sample by NEOs. We consider detections with a photometric uncertainty $\sigma < 0.05$ mag. A maximum change of $\Delta m < 1.2$ mag was allowed between detections as it would be unphysical for an asteroid to show variations greater than this due to rotation on a 15 minute timescale. Using both the time between detections and the difference in brightness the absolute rate of change in magnitude was calculated for each detection pair within a tracklet.

The dataset was filtered down to only include objects which showed 3 or more values of rate of change in magnitude $|\dot{m}| > 0.3$ mag per 15 minutes spread across 2 or more tracklets. We also favoured objects which showed similar variations in the later data (May 2012 onwards). This gave a list of 38 potentially interesting targets for follow-up. In order to further reduce the likelihood that the large variations for these objects were caused by bad pixels or background contamination each image for the 38 targets was manually inspected. Using this  method it was found that 5 of the objects had changes in brightness that were caused by mechanisms other than rotation and hence these were removed from the target list. 

The target list was divided into primary and secondary targets according to the largest observed change in brightness. Primary targets were defined as those showing absolute rates of change in magnitude $|\dot{m}| > 0.6$ mag per 15 minutes. There were 10 primary targets defined this way. All remaining objects were named secondary targets. 

\section{Observations}

In the course of this project 22 nights were spent observing on the Isaac Newton Telescope and the New Technology Telescope. We were supported by follow-up observations throughout the duration of the project using the University of Hawaii 2.2m telescope.

\subsection{Isaac Newton Telescope}
\label{subsec:haa_int}

The 2.5m Isaac Newton Telescope (INT) is located at the Roque de los Muchachos Observatory on La Palma in the Canary Islands where it is a part of the Isaac Newton Group. The instrument used for the observations within this project was the Wide Field Camera (WFC). The WFC is a four chip mosaic comprised of EEV 2048x4100 pixel CCDs. The CCDs have a pixel size of 13.5 microns giving a pixel scale of 0.33" per pixel. Images taken using the INT are recorded as multi-extension fits files with an extension for each of the four CCDs. For the purposes of this investigation we were only concerned with the data on the central CCD, CCD4. 

Our initial observing strategy was to obtain at least 3.3 hours of coverage on a single object, allowing for at least 1.5 rotations of a super fast rotator. These observations were carried out using the Sloan $r$ filter with fast read-out (a read out time of 29 seconds) and over an exposure time selected to ensure the object's motion did not exceed 0.5". In poor seeing we increased this exposure time to ensure we obtained the required signal-to-noise. We later adjusted this observing strategy to observe multiple targets in order to obtain maximum coverage of our target list.  For this method we selected 2 or 3 visible targets and cycled between them in intervals of 10-20 minutes.

\subsection{New Technology Telescope}
\label{subsec:haa_ntt}

The 3.58m ESO New Technology Telescope (NTT) is located at La Silla in the Atacama Desert, Chile. The NTT is a Ritchey-Chr\'{e}tien telescope on an altazimuthal mount. The instruments on the NTT are mounted on its two Nasmyth foci. The instrument used in this investigation was the ESO Faint Object Spectrograph and Camera (EFOSC2). The EFOSC2 CCD is 2048x2048 and the field of view is 4.1'x4.1' with a pixel scale of 0.12" per pixel. The observations presented here were taken using the Sloan r filter and with 'normal' readout speed (40 seconds). For this we used 2x2 binning.

As with the INT observations exposure times were calculated for each object to obtain maximum signal-to-noise while keeping the asteroid's motion in that interval to $<0.5$". The same observing strategies were used for the NTT observations as for the INT.

\section{Analysis}
\label{sec:analysis}

To search for periodicities in the brightness variations of the observed asteorids we use the Lomb-Scargle method to generate periodograms. We favour the Lomb-Scargle method in this case as we have irregularly spaced data, which this method can easily handle unlike e.g. fitting a Fourier model which would require regularly spaced data.


The main drawback of the original LS periodogram is that it does not account for uncertainties in the input data. As we are dealing primarily with high-quality photometric data measured from high amplitude objects this is unlikely to affect our result too much. However, we still elected to use the 'Generalised Lomb-Scargle Periodogram' (\citealt{zechmeister}) which takes these uncertainties into account. This method also fits for the mean of the observed data rather than simply assuming that it is identical to that of the fitted sine curve, an assumption made by the initial form of the LS periodogram.







To estimate the uncertainty in the period obtained we use Equation~\ref{eqn:uncertain} (\citealt{horne1986}). In this equation A is the lightcurve amplitude, T is the total time spread of the dataset and $\sigma_{N}$ is the variation of the noise in the data. Assuming that the variation in the asteroid lightcurves is due to rotation of the object we assume the most likely period to be twice the best fit period from our Lomb-Scargle periodogram. This will produce the expected double peaked lightcurve from rotation.

\begin{equation} 
\delta\omega=\frac{3\pi\sigma}{2\sqrt{N}TA}
\label{eqn:uncertain}
\end{equation}

For each object we consider the following cases as potential shape solutions: a single ellipsoidal rubble pile object and a binary or bilobed approximation. For the case of a single elongated object we use Equations~\ref{eqn:chandraparty} and \ref{eqn:Delta} to obtain a Jacobi ellipsoid shape solution for the axis ratios of the body. A density for this strengthless solution can then be obtained using Equation~\ref{eqn:density2}. In addition to the uncertainty in this value due to the uncertainties in the shape and spin properties of the object we also apply an uncertainty due to internal friction. Following the method of \cite{holsapple2001} we obtain a range of densities for which the Jacobi shape solution applies assuming an angle of friction $\phi_{F} \le 15^{\circ}$. The required strength of an object in these uncertainty ranges can then be determined using Equations~\ref{eqn:sigmax} - \ref{eqn:slope}. This strength value provides a sensible upper limit on the internal strength of these objects as a lower angle of friction than this in an asteroid is unlikely.

For the case of a bilobed or contact binary as detailed in Sections 1.3 and 1.4 we use the model of \cite{lacerda2007} to obtain the shape and density for this configuration. We also obtain solutions where possible for bilobed objects along the bilobed equilibrium sequence as defined by \cite{descamps2015}.

In cases where there is a clear unambiguous solution for a single elongated object we favour this shape configuration. For those objects where the solution for a single elongated object requires unusual density or strength we also consider binary or bilobed solutions to be valid. It should be emphasised, however, that from light curve data alone any binary or bilobed solutions should be considered speculative.

\section{Photometry}
\label{sec:haa_phot}

Data reduction and photometry were performed using IRAF software package $apphot$ (\citealt{davis1999}). For each night of observation at least 30 bias frames were taken and combined into a single average frame for bias subtraction. Each night at twilight sky exposures were taken and combined and normalised to generate a flat field image.

In this work we use differential photometry rather than absolute photometry. For these observations, however, we had access to the stellar magnitudes catalogued by the PS1 $3\pi$ survey which measured the magnitudes of all stars in its sensitivity range above declination $-30^{\circ}$. This survey data was part of the first public data release in late 2016 (\citealt{magnier2016}; \citealt{flewelling2016}). With this data differential photometry becomes straightforward and allows us to obtain accurate measurements even in non-photometric conditions. 

Due to the relatively fast motion of the asteroids observed, it was impossible to use the same stars on multiple nights so instead it was necessary to select suitable different stars for each night. At least 5 stars were used in each set of observations to minimise the risk of contamination by stellar phenomena e.g. variability.  To ensure none of our background stars will affect our data adversely we plot the difference in brightness between each combination of two background stars and remove any stars showing significant variation with time. Any other factors acting on the brightness of both the asteroid and background stars e.g. cloud cover or atmospheric effects, can be assumed to affect all bodies equally.

Measurements were taken with circular apertures centred around both the target asteroid and the background stars. The radius of this aperture was chosen based on the FWHM of the background stars in pixels to maximise the obtained signal-to-noise (\citealt{howell1989}). For all targets observed at declination $> -30^{\circ}$ we are able to make use of PanSTARRS $3\pi$ survey data to obtain accurate magnitudes for these background stars, for those targets observed with the NTT at declination $< -30^{\circ}$ we made use of the Gemini South GMOS-S Photometric Standard Star Fields.

Light curve amplitude is dependent on the phase angle of the observations. Scattering effects and increased shadowing at high phase angles will cause light curve minima to appear fainter and hence causes the light curve amplitude to be increased. This can lead to an overestimation of the elongation of an object. \cite{zappala1990} demonstrated a linear relationship between the apparent amplitude of a light curve $A_{obs}$ and its actual amplitude $A(\alpha=0^{\circ})$ for phase angles $\alpha \leq 40^{\circ}$. This is given in Equation~\ref{eqn:phase} where $s$ is a taxonomy-dependent slope parameter.

\begin{equation}
A(\alpha=0^{\circ}) = \frac{A_{obs}}{1+s\alpha}
\label{eqn:phase}
\end{equation}

All reported light curve observations in this project were made in the main asteroid belt and hence fall within this $\alpha$ domain allowing Equation~\ref{eqn:phase} to be used. As the photometric data obtained as part of this project alone is not enough to determine the taxonomy of any of our target objects we use a conservative value of $s=0.015$ mag deg$^{-1}$ in each case.

\section{Observational Results}

A summary of the orbital and observational properties of each asteroid discussed in this section is presented in Tables \ref{table:summary}-\ref{table:rotsummary}.

\begin{table}
\caption{Table of Observations} 
\centering 
\begin{tabular}{c c c c c c c c c c} 
\hline\hline 
Object & Date (UT) & Obs. Hours & Images & Filter & Telescope & $r (AU)$ & $\Delta$ (AU) & $m_{V}$ & $\alpha$ (deg) \\  
\hline 
39684 & 29th April 2015 &2.7& 85 & r & INT & 3.08 & 2.13 & 18.83 & 8\\
39684 & 30th April 2015 &3.8& 138 & r & INT & 3.08 & 2.13 & 18.81 & 7\\
39684 & 1st May 2015 &4.6& 60 & r & INT & 3.08 & 2.12 & 18.79 & 7\\
39684 & 2nd May 2015 &4.7& 142 & R & UH2.2m & 3.08 & 2.11 & 18.77 & 7\\
\hline 
45864 & 22nd January 2015 &2.0& 107 & R & UH2.2m & 2.16 & 1.83 & 18.30 & 27\\
45864 & 27th January 2015 &2.6& 127 & R & UH2.2m & 2.15 & 1.90 & 18.38 & 27\\
45864 & 28th January 2015 &2.9& 144 & R & UH2.2m & 2.15 & 1.91 & 18.39 & 27\\
45864 & 15th January 2016 &1.9& 104 & R & UH2.2m & 2.20 & 2.06 & 18.57 & 26\\
45864 & 18th January 2016 &1.7& 120 & R & UH2.2m & 2.20 & 2.03 & 18.54 & 27\\
45864 & 15th April 2016 &1.4& 41 &r& INT & 2.30 & 1.31 & 16.94 & 5\\
45864 & 16th April 2016 &1.5& 20 & r& INT & 2.30 & 1.31 & 16.92 & 5\\
45864 & 6th May 2016 &4.2& 23 &r&UH2.2m & 2.33 & 1.35 & 17.16 & 8\\
45864 & 27th May 2016 &2.7& 35 &r&INT & 2.35 & 1.49 & 17.71 & 17\\
\hline 
45721 & 29th April 2015 &3.4& 96 & r & INT & 4.04 & 3.07 & 19.03 & 4\\
45721 & 30th April 2015 &5.5& 81 & r & INT & 4.04 & 3.07 & 19.04 & 4\\
\hline 
18018 & 1st May 2015 &4.3& 79 & r & INT & 2.19 & 1.45 & 18.08 & 22\\
18018 & 6th May 2015 &3.0& 106 & r & WHT & 2.19 & 1.50 & 18.19 & 23\\
18018 & 9th June 2015 &1.4& 55 & R & UH2.2m & 2.21 & 1.90 & 18.83 & 27\\
18018 & 10th June 2015 &1.3& 52 & R & UH2.2m & 2.21 & 1.91 & 18.85 & 27\\
\hline 
18280 & 16th January 2016 &1.1& 19 & r & INT & 2.12 & 1.35 & 17.50 & 21\\
18280 & 16th January 2016 &1.8& 63 & R & UH2.2m & 2.12 & 1.35 & 17.50 & 21\\
18280 & 18th January 2016 &0.5& 15 & r & INT & 2.12 & 1.36 & 17.55 & 21\\
18280 & 19th January 2016 &2.9& 191 & R & UH2.2m & 2.12 & 1.37 & 17.57 & 22\\
\hline 
206167 & 18th July 2015 &4.5& 57 & r & NTT & 2.70 & 1.83 & 20.26 & 14\\
206167 & 23rd July 2015 &2.1& 54 & R & UH2.2m & 2.69 & 1.87 & 20.35 & 15\\
206167 & 24th July 2015 &2.3& 77 & R & UH2.2m & 2.69 & 1.88 & 20.37 & 16\\
\hline 
49257 & 17th July 2015 &2.6& 108 & r & NTT & 2.36 & 2.17 & 19.62 & 26\\
49257 & 18th July 2015 &2.7& 120 & r & NTT & 2.36 & 2.16 & 19.61 & 26\\
49257 & 23rd July 2015 &1.7& 58 & R & UH2.2m & 2.36 & 2.10 & 19.54 & 26\\
49257 & 18th September 2015 &4.3& 172 & R & UH2.2m & 2.35 & 1.50 & 18.50 & 16\\
\hline 
\end{tabular}
\label{table:summary} 
\end{table}

\clearpage

\begin{table}
\caption{Table of Orbital and Physical Properties of High Amplitude Asteroids} 
\centering 
\begin{tabular}{c c c c c c c} 
\hline\hline 
Object & a (AU) & e & i ($^{\circ}$) & H & Albedo & D (km)  \\  
\hline 
$39684$ & $2.92$ & $0.07$ & $2.89$ & $14.2$ & $0.12\pm 0.04^{1}$ & $4.5 \pm 1.0$\\
$45864$ & $2.31$ & $0.10$ & $5.89$ & $14.1$ & $0.35\pm 0.05^{1}$ & $4.3 \pm 0.2$\\
$45721$ & $3.80$ & $0.07$ & $1.02$ & $13.2$ & $0.10\pm 0.01^{1}$ & $8.5 \pm 0.2$\\
$18018$ & $2.20$ & $0.05$ & $6.19$ & $14.5$ & $0.30\pm 0.08^{1}$ & $3.3 \pm 0.4$\\
$18280$ & $2.42$ & $0.13$ & $7.54$ & $14.3$ & $0.21\pm 0.10^{2}$ & $4.5 \pm 1.2$\\
$206167$ & $2.78$ & $0.17$ & $2.72$ & $16.0$ & $0.17\pm 0.08^{2}$ & $2.3 \pm 0.6$\\
$15613$ & $2.57$ & $0.11$ & $12.8$ & $13.9$ & $0.23\pm 0.03^{1}$ & $6.0 \pm 0.1$\\
$49257$ & $2.62$ & $0.10$ & $5.66$ & $14.9$ & $0.27\pm 0.08^{1}$ & $3.1 \pm 0.3$\\
\hline 
\end{tabular}
\\
$^{1}$ WISE/NEOWISE albedos (\citealt{Masiero2011})\\
$^{2}$ Average albedo for 0.5 AU annuli of main asteroid belt (\citealt{mcneill2016})
\label{table:orbsummary} 
\end{table}

\begin{table}
\caption{Summary Table of Rotational Properties of High Amplitude Asteroids} 
\centering 
\begin{tabular}{c c c c} 
\hline\hline 
Object & P (h) & A (mag) &  A$_{0^{\circ}}$ (mag)\\  
\hline 
$39684$ & $4.836 \pm 0.016$ & $1.29 \pm 0.02$ &  $1.17 \pm 0.02$\\
$45864$ & $5.135 \pm 0.001$ & $1.61 \pm 0.01$ &  $1.15 \pm 0.01$\\
$45721$ & $6.00 \pm 0.06$ & $1.11 \pm 0.07$  & $1.05 \pm 0.07$\\
$18018$ & $5.608 \pm 0.002$ & $1.33 \pm 0.03$  & $1.00 \pm 0.03$\\
$18280$ & $3.808 \pm 0.008$ & $1.15 \pm 0.05$ & $0.87 \pm 0.05$\\
$206167$ & $3.944 \pm 0.002$ & $1.05 \pm 0.15$ & $0.86 \pm 0.12$\\
$15613$ & $3.3 \pm 0.4$ & $1.02 \pm 0.02$ & $0.78 \pm 0.02$\\
$49257$ & $3.3896 \pm 0.0002$ & $0.80 \pm 0.05$ & $0.58 \pm 0.05$\\
\hline 
\end{tabular}
\label{table:rotsummary} 
\end{table}

\subsection{39684 (1996 PD8)}
\label{sub:haa_39684}



\begin{figure}[H]
  \begin{center}
\includegraphics[width=0.5\textwidth]{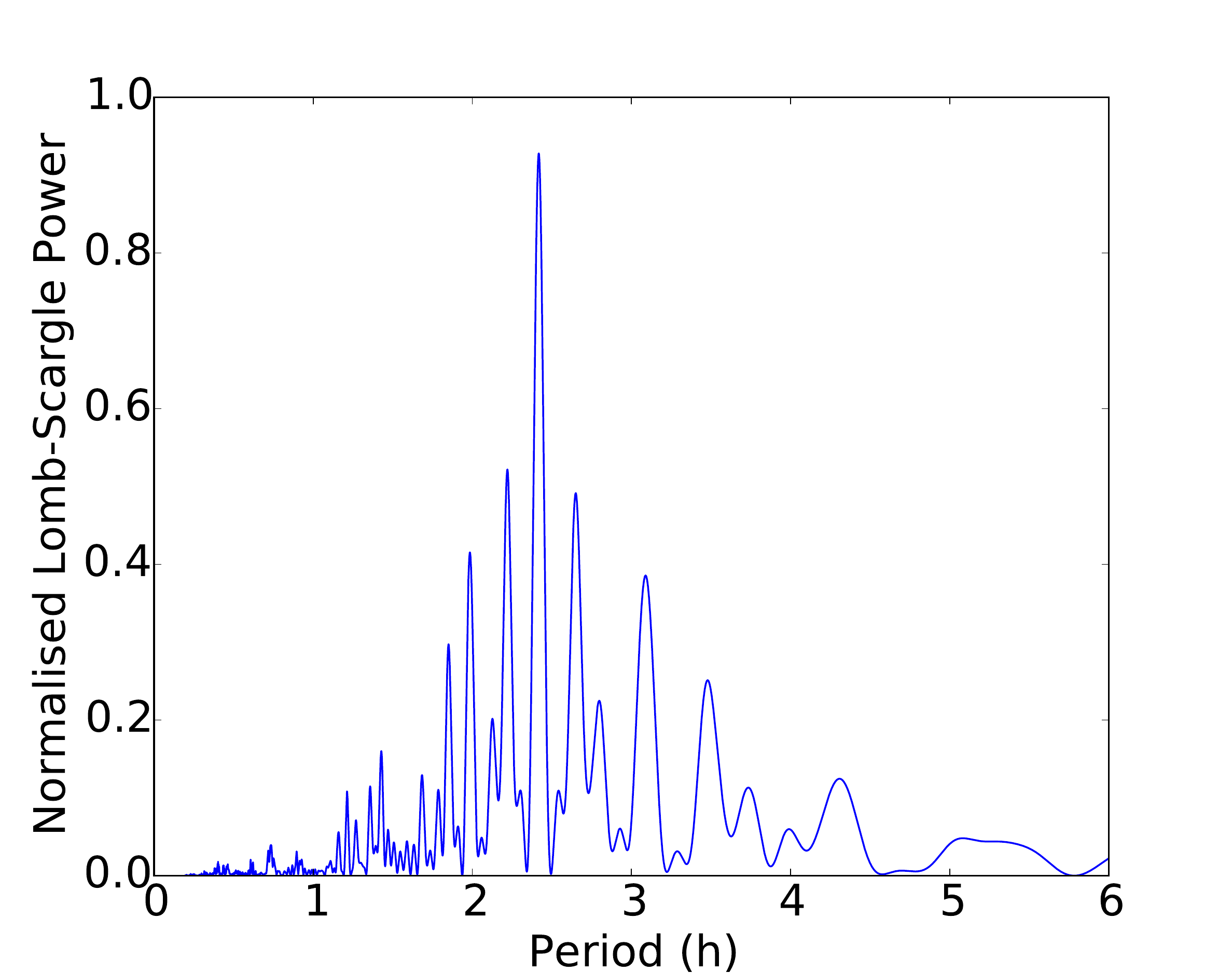}
   \caption{The Lomb-Scargle periodogram generated from the light curve data of 39684 }
\label{fig:39684_ls}
  \end{center}
 \end{figure}

The Lomb-Scargle periodogram calculated from these observations is given as Figure~\ref{fig:39684_ls}.


\begin{figure}
  \begin{center}
\includegraphics[width=0.5\textwidth]{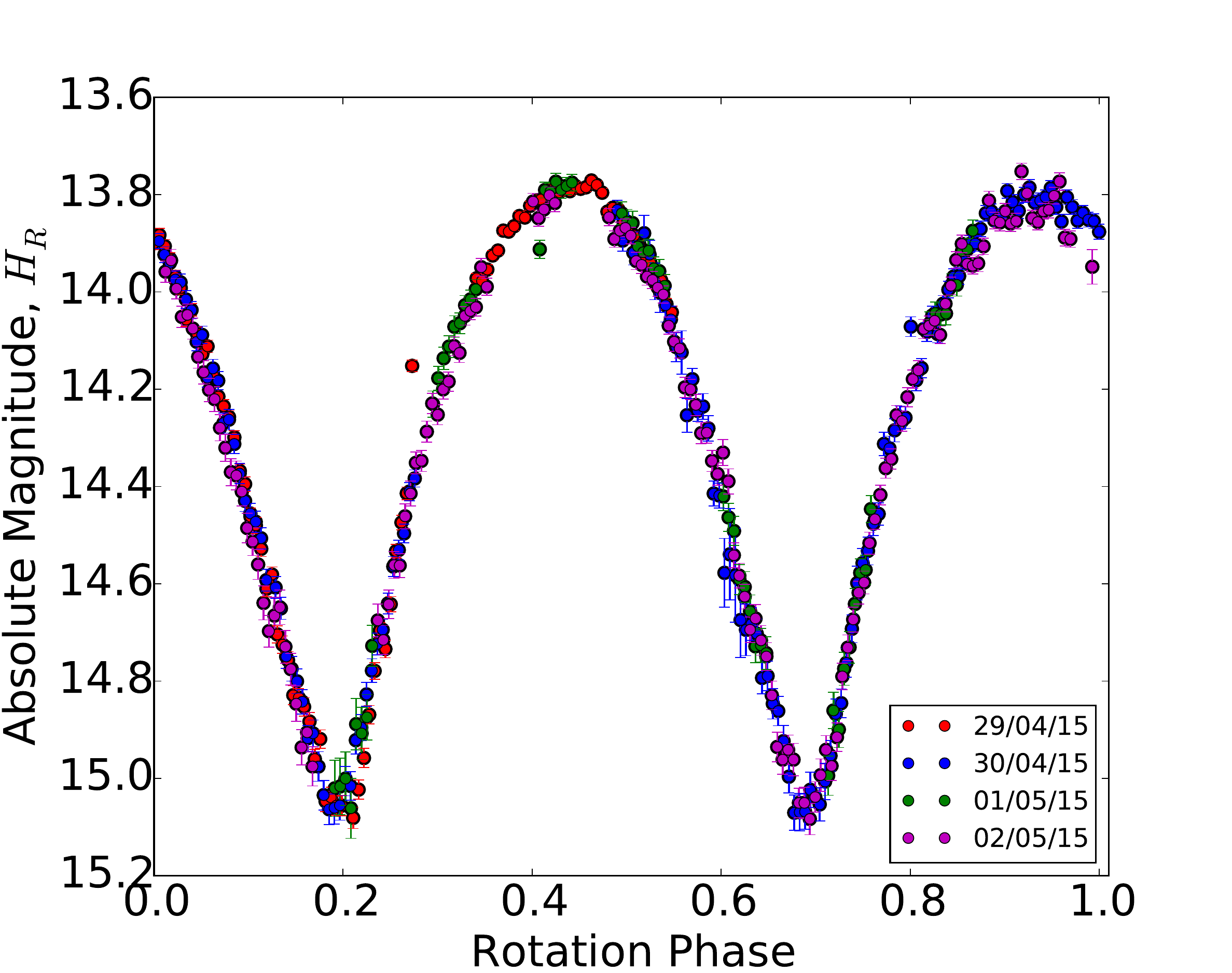}
   \caption{A light curve constructed from photometric data from observations of asteroid 39684 using the Isaac Newton Telescope in April and May 2015 and phased to the determined rotation period $P_{r}=4.836\pm 0.016$ h.}
\label{fig:39684}
  \end{center}
 \end{figure}

The periodogram finds a best fit rotation period of $P_{r}=4.836\pm 0.016$ h. The measured light curve folded to this rotation period showed an amplitude of $1.29\pm 0.02$ magnitudes and is given in Figure~\ref{fig:39684}. Accounting for phase angle this amplitude scales to $1.17\pm 0.02$ mag.

For these values we find a best result for a Jacobi ellipsoid with axis ratios $1:0.34 \pm 0.01 :0.29 \pm 0.01$ and density $\rho=2600 \pm 50$ \kgm. The elongation of this object would imply that if it were a rubble pile that it would be unstable and mass loss would occur. For this object to be stable significant cohesive strength would be required. From the work of \cite{holsapple2001} we find a range of densities allowed for an ellipsoid of this shape assuming $\phi=15^{\circ}$ of $1900<\rho<5900$ \kgm. 


The Roche model produced a lower limit fit for this object of density $\rho=4400\pm 200$ \kgm. Considering the porosity required for Roche deformation this is an extremely high density value approaching the maximum we defined for asteroids for porosity $\sim 55\%$ . This density seems implausible for the case of a Roche tidally-distorted bilobed object, although we cannot entirely rule it out. 

For objects along the bilobed equilibrium sequence rotating with $P_{r}=4.840\pm 0.006$ h we obtain density values $2900<\rho<5900$ \kgm across the range of $\Omega$ values. Here, the only solutions that give densities that could be plausible are those close to the Jacobi bifurcation ellipsoid i.e. $c' \approx c$. These solutions may produce the observed light curve and we must consider them a possible solution.

In summary we find the most viable solution is obtained for a triaxial ellipsoid potentially with some degree of cohesive strength. We find potential solutions for objects with density $1900<\rho<5900$ \kgm, axis ratio $1:0.34 \pm 0.01:0.29 \pm 0.01$ and internal cohesive strength $50-530$ Pa. There are some solutions on the bilobed equilibrium sequence that also may explain this asteroid's light curve however there is no evidence at this time to suggest that this is a more likely solution than a simple ellipsoid. 


\subsection{45864 (2000 UO97)}
\label{sub:haa_45864}



\begin{figure}
  \begin{center}
\includegraphics[width=0.5\textwidth]{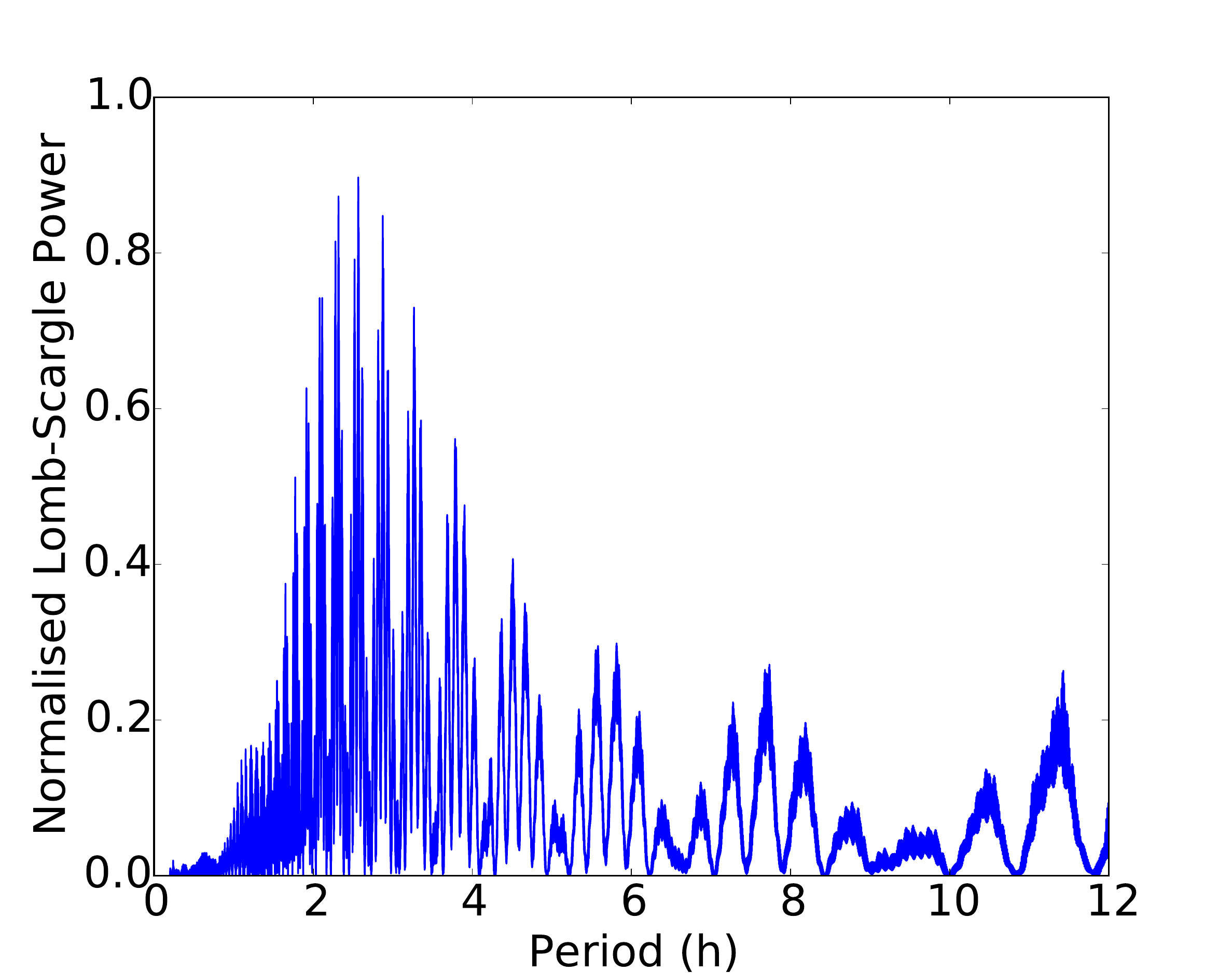}
   \caption{The Lomb-Scargle periodogram generated from the light curve data of 45864 from January 2015.}
\label{fig:45864_ls}
  \end{center}
 \end{figure}

This object represents the most well-observed target from our initial list. The generalised Lomb-Scargle periodogram for this is given in Figure~\ref{fig:45864_ls}.


\begin{figure}
  \begin{center}
\includegraphics[width=0.5\textwidth]{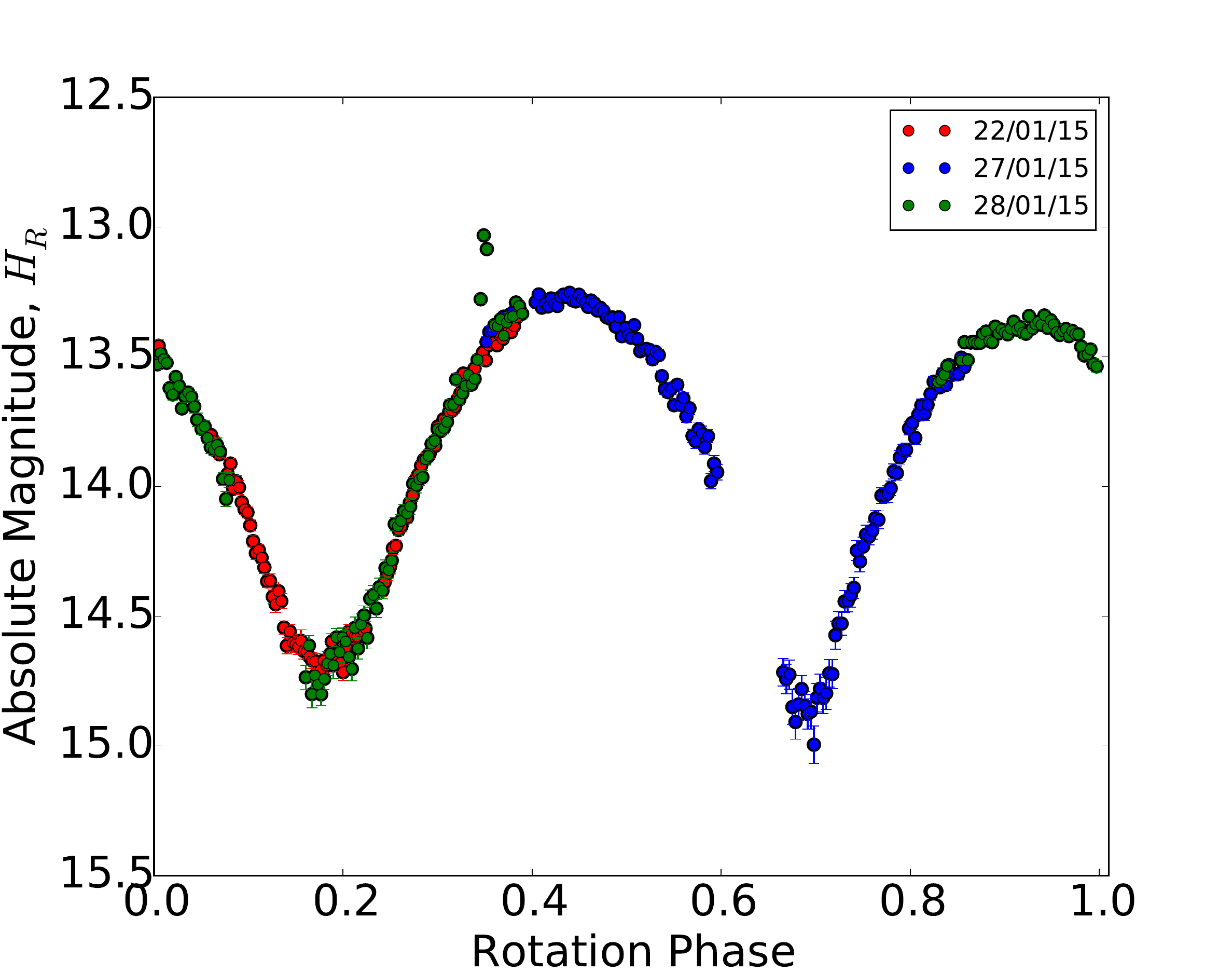}
   \caption{A light curve constructed from photometric data from observations of asteroid 45864 using the University of Hawaii 2.2m Telescope in January 2015 and phased to the determined rotation period $P_{r}=5.135$ h.}
\label{fig:45864_lc1}
  \end{center}
 \end{figure}

\begin{figure}
  \begin{center}
\includegraphics[width=0.5\textwidth]{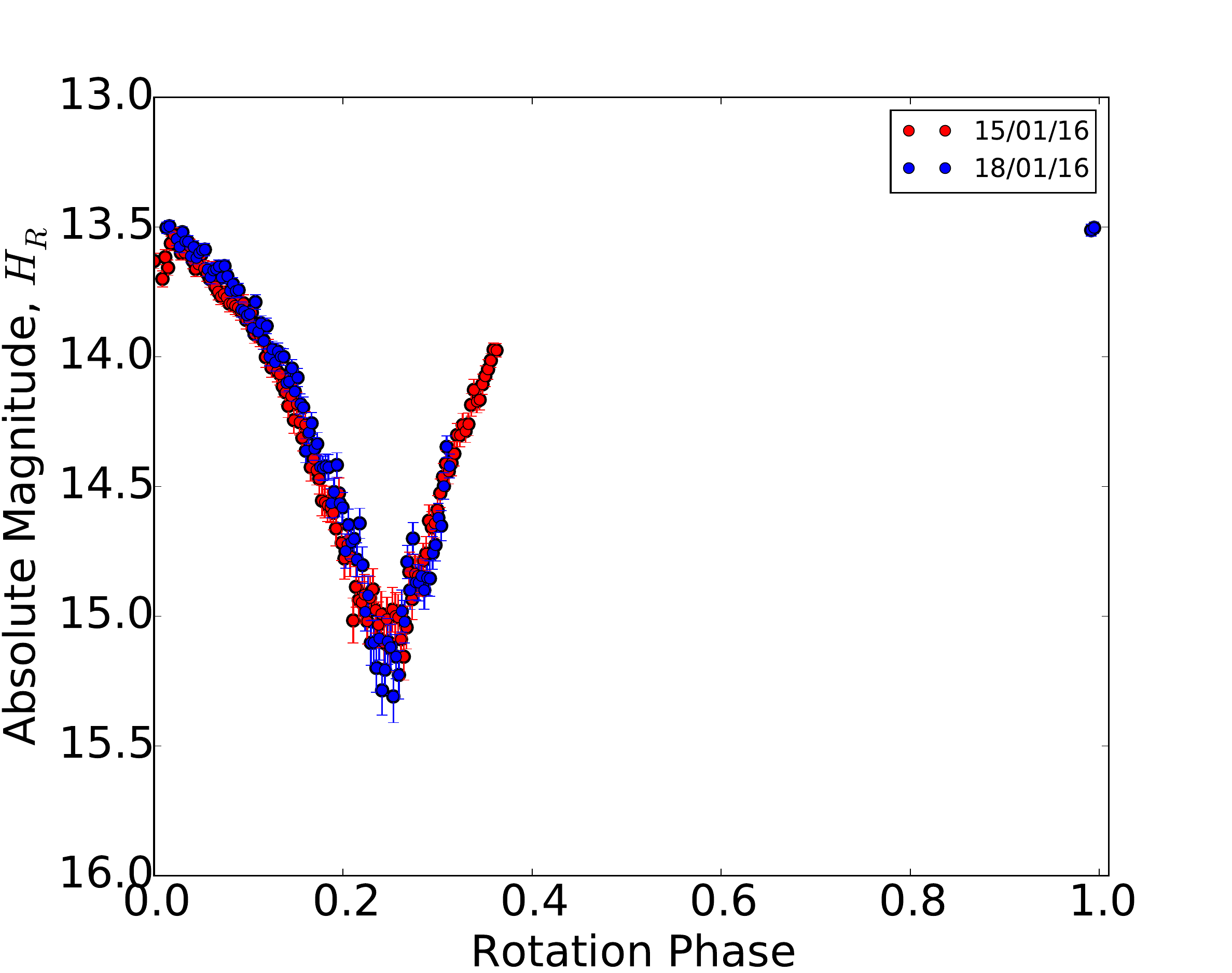}
   \caption{A light curve constructed from photometric data from observations of asteroid 45864 using the University of Hawaii 2.2m Telescope in January 2016 }
\label{fig:45864_lc2}
  \end{center}
 \end{figure}

\begin{figure}
  \begin{center}
\includegraphics[width=0.5\textwidth]{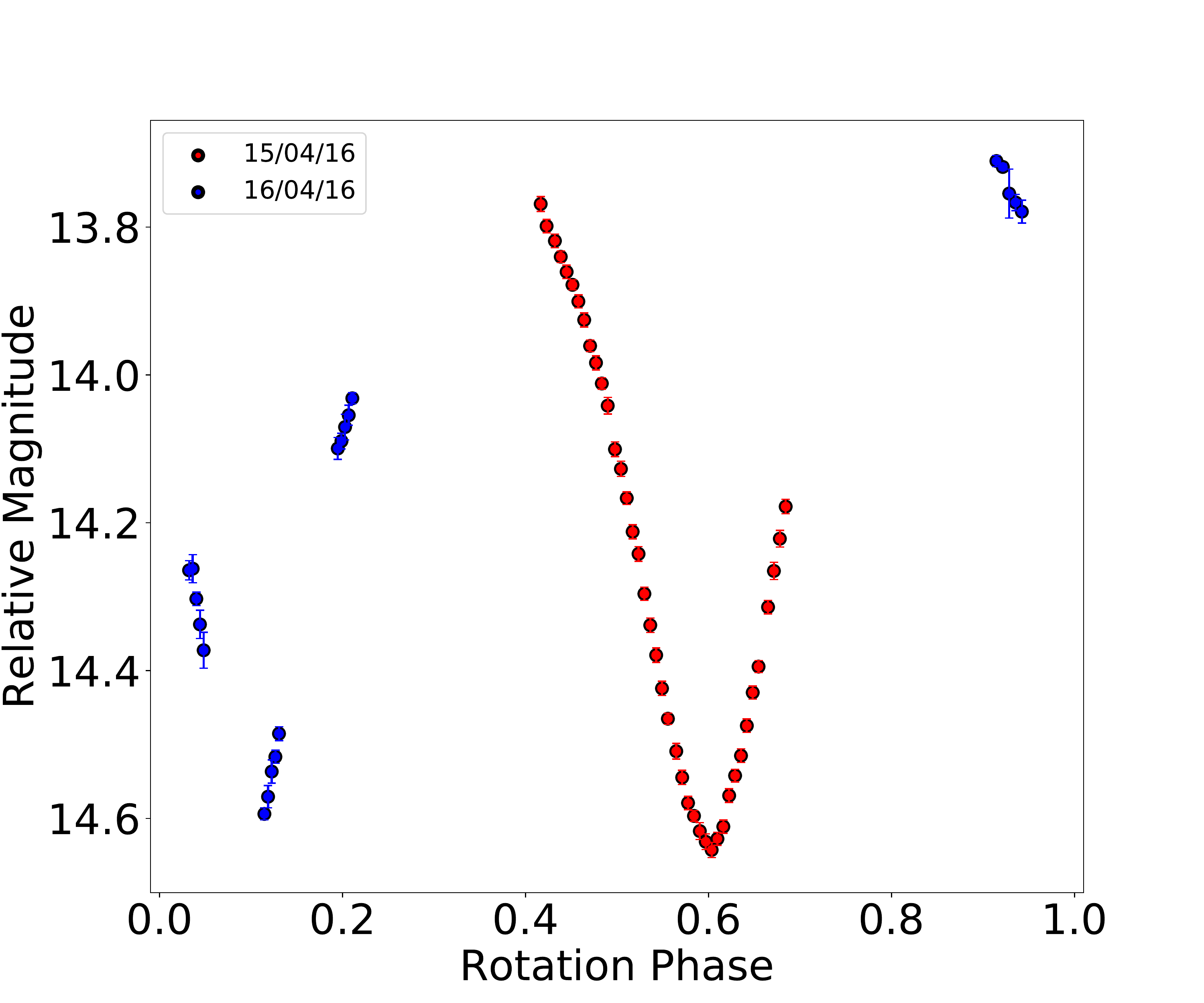}
   \caption{A light curve constructed from photometric data from observations of asteroid 45864 using the Isaac Newton Telescope in April 2016 }
\label{fig:45864_lc4}
  \end{center}
 \end{figure}

\begin{figure}
  \begin{center}
\includegraphics[width=0.5\textwidth]{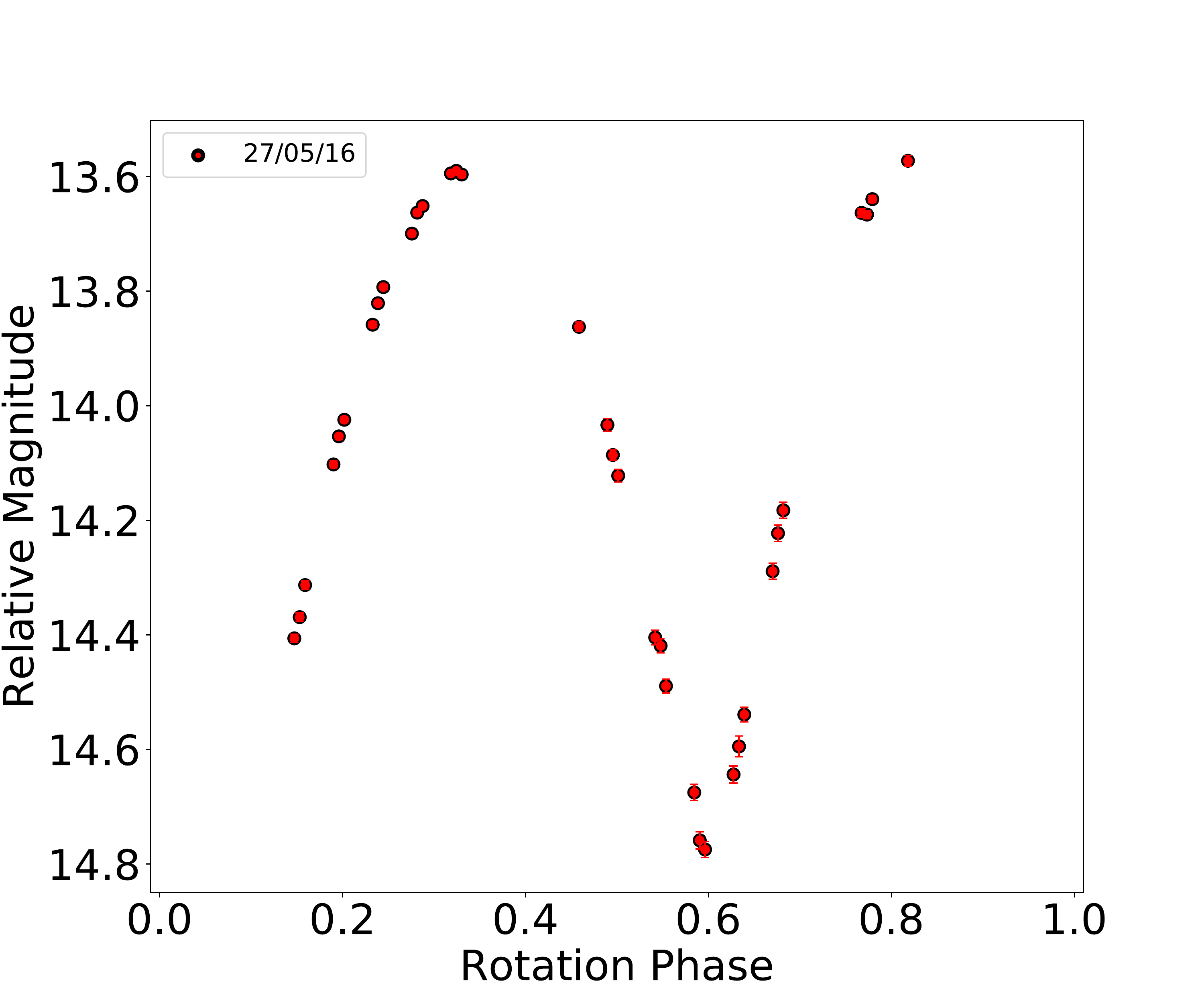}
   \caption{A light curve constructed from photometric data from observations of asteroid 45864 using the University of Hawaii 2.2m Telescope in May 2016 }
\label{fig:45864_lc3}
  \end{center}
 \end{figure}

Figure~\ref{fig:45864_ls} shows a best fit rotation period of $P_{r}=5.135\pm 0.001$ h. Light curves folded to this period from the data are given in Figures~\ref{fig:45864_lc1}-\ref{fig:45864_lc3}.

Since these observations were carried out \cite{durech2016} calculated a period for this object from sparse archival data and obtained a value of $5.13544 \pm 0.00001$ h, in agreement to within uncertainties of our observed value. 


The maximum amplitude observed for this object was found to be $1.61\pm 0.01$ magnitudes, an extremely large variation for a main belt asteroid of this size ($\sim4.3$ km). Adjusting for the relatively high phase angle of these observations the real amplitude is closer to $1.15 \pm 0.01$ magnitudes.  In Figures~\ref{fig:45864_lc4} and ~\ref{fig:45864_lc3} the corrected amplitudes of the light curve were found to be $0.93$ and $0.97$ magnitudes, respectively. This change is potentially due to the change in orbital geometry between the sets of observations.

For an object with this rotation rate and showing this magnitude variation we find that for this to be explained by a Jacobi ellipsoid would require an object with axis ratios $1:0.35 \pm 0.01 :0.29 \pm 0.01$ and a bulk density $\rho=2300 \pm 50$ \kgm. The albedo of this object is too high ($\sim0.35$) a value for the most abundant S and C classes of asteroids (\citealt{mainzer2011}). We consider the possibility that this object is instead an E-type enstatite chondritic asteroid.  If this object were a rubble pile Jacobi ellipsoid the axis ratios required for this object mean that the object is much too elongated to exist as a single object without undergoing mass loss, suggesting that the object would have to be monolithic. The internal friction acting between the solid aggregate of a rubble pile allows it to withstand break-up to a greater extent than a fluid body, although its strength will still be considerably less than that of a coherent body. 


\cite{holsapple2001} and \cite{holsapple2004} derived limits on the equilibrium configurations for an ellipsoid with no tensile strength, modelling them as elastic-plastic materials using a Mohr-Coulomb yield model dependent on angle of friction, $\phi_{F}$. \cite{sharma2009} showed that for most large asteroids the angle of friction $\phi_{F}<5^{\circ}$. Taking the Jacobi ellipsoid shape of $1:0.35:0.29$ and using the scaled spin rate ($\Omega=\frac{\omega}{\sqrt{G\rho}}$) vs axis ratio ($b/a$) plots of \cite{holsapple2001} there is no equilibrium figure for an object of this shape rotating with this rotation period where $\phi_{F}<5^{\circ}$. \cite{holsapple2001} shows that all asteroids put into their model have $\phi_{F}<40^{\circ}$ with most having $\phi_{F}<15^{\circ}$. Taking this boundary of $\phi_{F}<15^{\circ}$, we find a range of densities that an ellipsoid figure of equilibrium can take for an object with 45864's rotation period and obtained axis ratios. The range obtained was $1600<\rho<5200$ \kgm. 





Using the Drucker-Prager yield model (Equation~\ref{eqn:drucker}) and the calculated densities, we constrained a cohesive strength in the range $0-130$ Pa (assuming an angle of friction $\phi_{F} = 15^{\circ}$). A strengthless solution of a Jacobi ellipsoid is present at around $2300$ \kgm as predicted earlier in this section. The lower end of this density range represents the most plausible values of bulk density for a rubble pile. 



The result obtained from the method described by \cite{lacerda2007} gave a lower limit density of $3900\pm 200$ \kgm. This is an extremely high density and is again difficult to reconcile with the theory of asteroids being rubble piles. At a porosity of $50\%$ this bulk density would require a grain density approximately that of iron. As the assumption that the object is a binary consisting of two Roche ellipsoids does not provide a believable solution we must consider alternative explanations. For this obtained density to be correct the object would have to be a bilobed object consisting of two solid components. As these bodies would not be deformed by tidal forces they would be closer to spherical components than Roche ellipsoids. Objects of this nature could not produce a light curve amplitude as great as the one observed due to shape alone and it would be unlikely that albedo variegation could make up this large difference.

For an equilibrium figure on the bilobed object sequence rotating with $P_{r}=5.135\pm 0.001$ h we find acceptable density solutions across the $\Omega$ range from $2600<\rho<5200$ \kgm. These lower $\rho$ values are potentially plausible densities for this asteroid and would imply a less bilobed shape i.e. $c' \approx c$. Consider the case of the bifurcation ellipsoid with axis ratios $1:0.30:0.27$. The amplitude of the light curve for an object of this shape is approximately $A=1.30$ mag. We also consider the shape on the sequence with the greatest elongation ($a/b$) with axis ratios $1:0.27:0.24$ and a waist axis ratio $c'=0.23$. Assuming this to have the projected area of a standard ellipsoid we estimate a light curve amplitude of this object $A_{obs}=1.41$ mag, with this value being greater than the actual amplitude that would be produced as we neglect the decrease in projected area of the object caused by its waist. From these estimates a particularly elongated bilobed shape gives a valid solution for this asteroid.

In summary we find that it has not been possible to find a simple shape solution for 45864. The large amplitude of the object along with its rotation period would imply that this object cannot exist as a single elongated rubble-pile ellipsoid and a best fit for a bilobed object using the Roche approximation produced an infeasible density. We are instead forced to consider that this simplification may not be sufficient to explain the structure of 45864. For this object we consider a single triaxial ellipsoid with axis $1:0.35 \pm 0.01:0.29 \pm 0.01$, density $1600<\rho<5200$ \kgm and minimal internal strength in the range $0-130$ Pa to be the most valid solution. We hope to obtain further observational data for this object.

\subsection{45721 (2000 GZ42)}
\label{sub:haa_45721}



\begin{figure}
  \begin{center}
\includegraphics[width=0.5\textwidth]{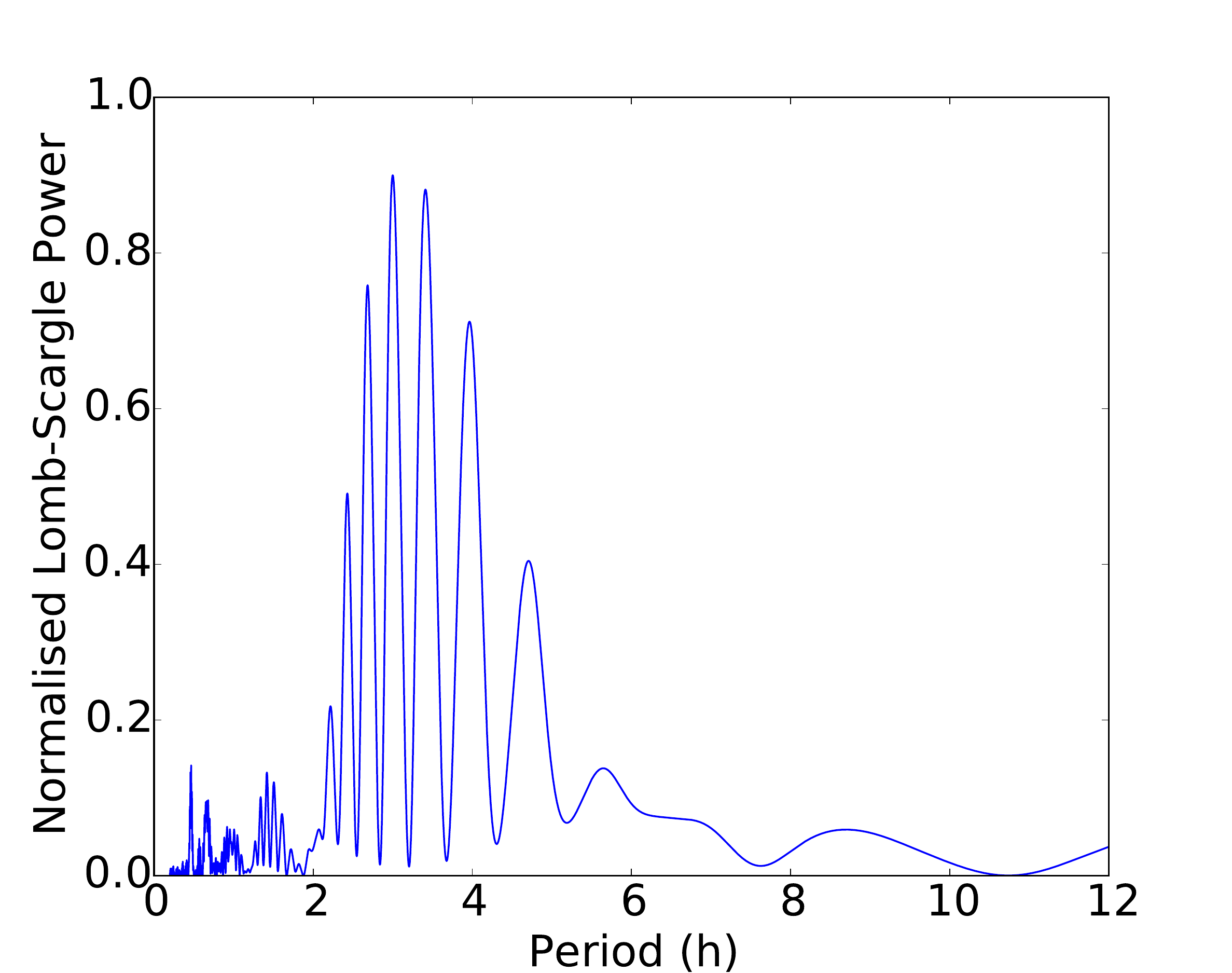}
   \caption{The Lomb-Scargle periodogram generated from the light curve data of 45721}
\label{fig:45721_ls}
  \end{center}
 \end{figure}

The high amplitude nature of this object was also observed by WISE and 45721 was identified as a potential target for observational follow-up as a potential close binary asteroid by both \cite{sonnett2014} and \cite{pedro}. The resulting LS periodogram is given as Figure~\ref{fig:45721_ls}.


\begin{figure}
  \begin{center}
\includegraphics[width=0.5\textwidth]{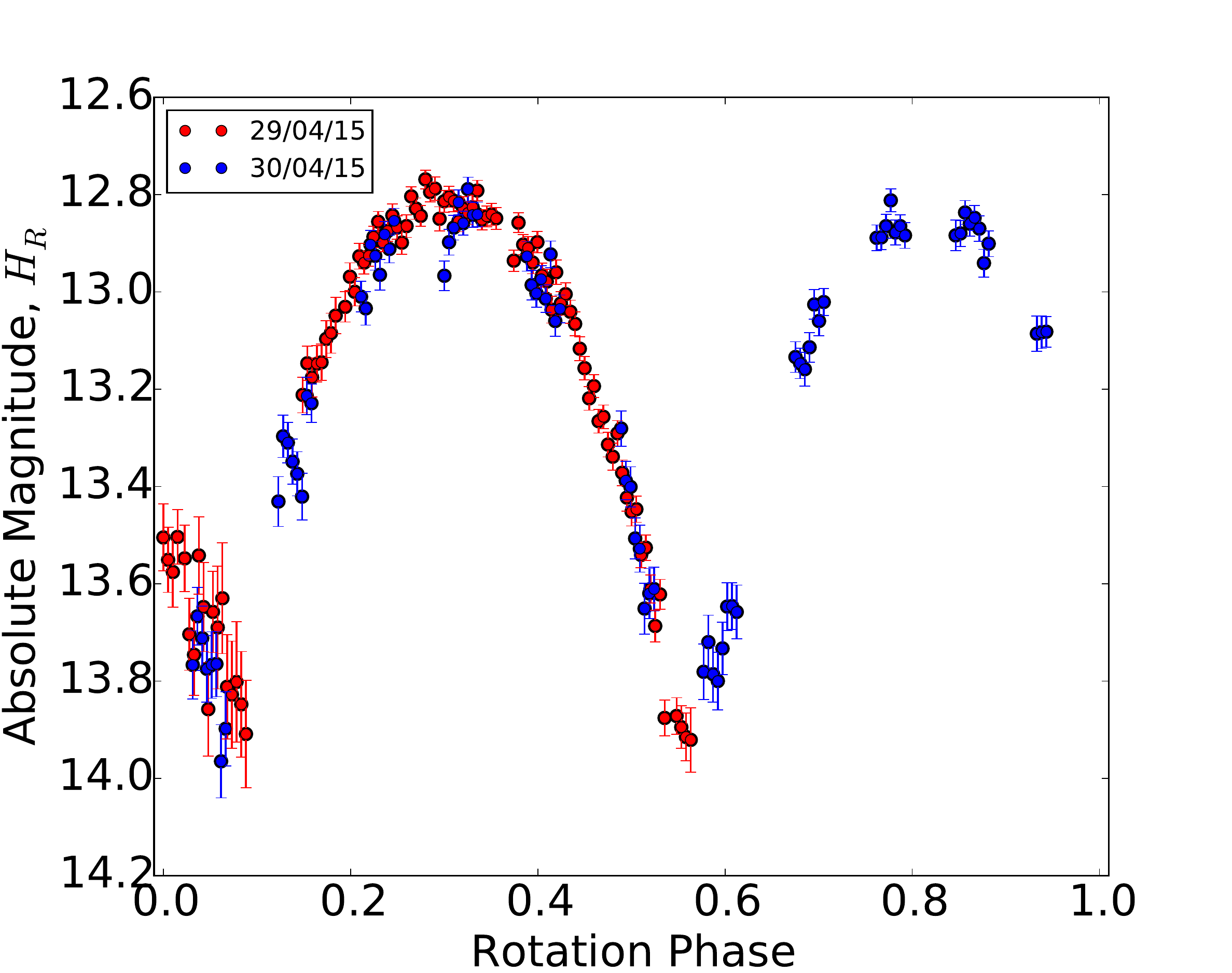}
   \caption{A light curve constructed from photometric data from observations of asteroid 45721 using the Isaac Newton Telescope in April 2015 and phased to the determined rotation period $P_{r}=6.00$ h.}
\label{fig:45721}
  \end{center}
 \end{figure}

The periodogram shows a best fit rotation period of $P_{r}=6.00 \pm 0.06$ h with a measured amplitude of $1.11\pm 0.07$ magnitudes. The phase angle of these observations was relatively small ($4^{\circ}$), and as such any correction to the amplitude is identical within uncertainties. The light curve phased with the best fit period is given in Figure~\ref{fig:45721}.

The Palomar Transient Factory identified this object as having a rotation period of $6.00 \pm 0.06$ h (\citealt{palomar}), in excellent agreement within $\sigma$ of the value that we obtained.


Assuming this object to be a Jacobi ellipsoid we find that for this spin period and amplitude an object with axis ratios $1:0.36 \pm 0.02:0.29 \pm 0.02$ and density $\rho=1600 \pm 100$ \kgm would be required. This is a plausible density for a rubble pile asteroid although this shape is close to the critical shape $b/a=0.43$ at which mass loss would be expected. Assuming an angle of friction $\phi_{F}=15^{\circ}$ we find a range of possible densities explaining this shape and spin rate $1100-4300$ \kgm. For this density range we then find that cohesive strengths of $0-120$ Pa provide the best solutions for cohesive strength across this range. 


From the Roche model for this object we find a density range $\rho=2800\pm 300$ \kgm, a plausible range of values for low-porosity chondritic asteroids. However, there is no clear evidence to suggest that a bilobed solution of this kind is any more likely than a single elongated shape solution.

Solving for density along the bilobed equilibrium sequence with rotation rate $P_{r}=6.00\pm 0.06$ h we obtain solutions $1900<\rho<3800$ \kgm. Where $c'$ approaches zero the density required would be too high to be plausible, exceeding that of solid chondritic chondrite ($3300$ \kgm) and potentially requiring low porosity metallic composition. However, for larger $c'$ values the densities are more in line with the known density of asteroids. While bilobed shape may be a plausible explanation for this object and we have to consider it as such, its amplitude $A_{obs}=1.1$ mag may also be more easily explained by a simpler shape and we have no specific evidence to suggest that this object is likely to be bilobed in nature.

In summary for this object we find that the observed light curve can be explained by a Jacobi ellipsoid with axis ratios $1:0.35 \pm 0.02:0.29 \pm 0.02$ and density $\rho=1600 \pm 100$ \kgm. Acceptable explanations with plausible density values may also be obtained for a Roche binary configuration or for a bilobed object. Further observations of this object in the future will be needed to draw any firmer conclusions.

\subsection{18018 (1999 JR125)}
\label{sub:haa_18018}



\begin{figure}
  \begin{center}
\includegraphics[width=0.5\textwidth]{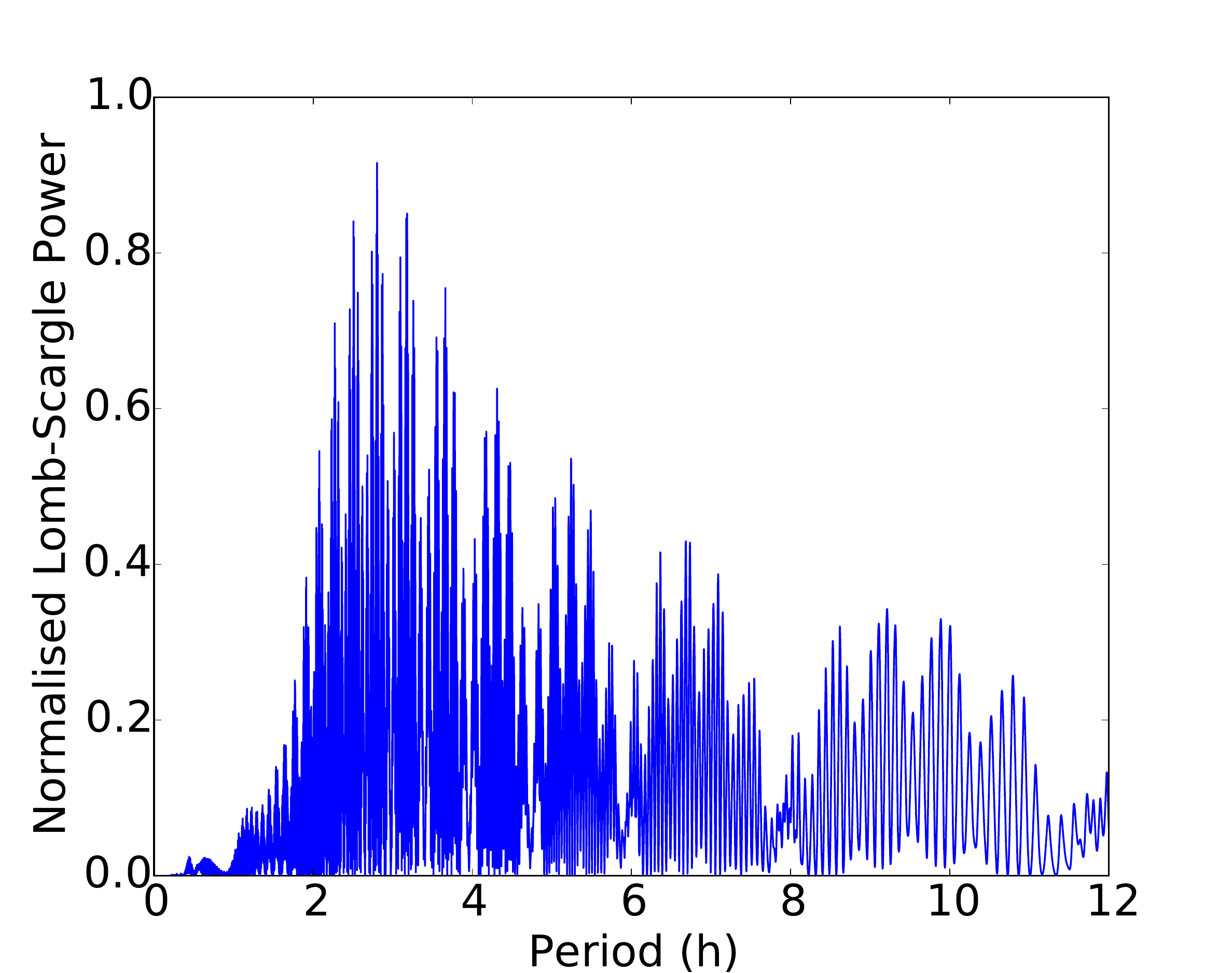}
   \caption{The Lomb-Scargle periodogram generated from the light curve data of 18018 }
\label{fig:18018_ls}
  \end{center}
 \end{figure}

The periodogram obtained from our data is presented in Figure~\ref{fig:18018_ls}.


\begin{figure}
  \begin{center}
\includegraphics[width=0.5\textwidth]{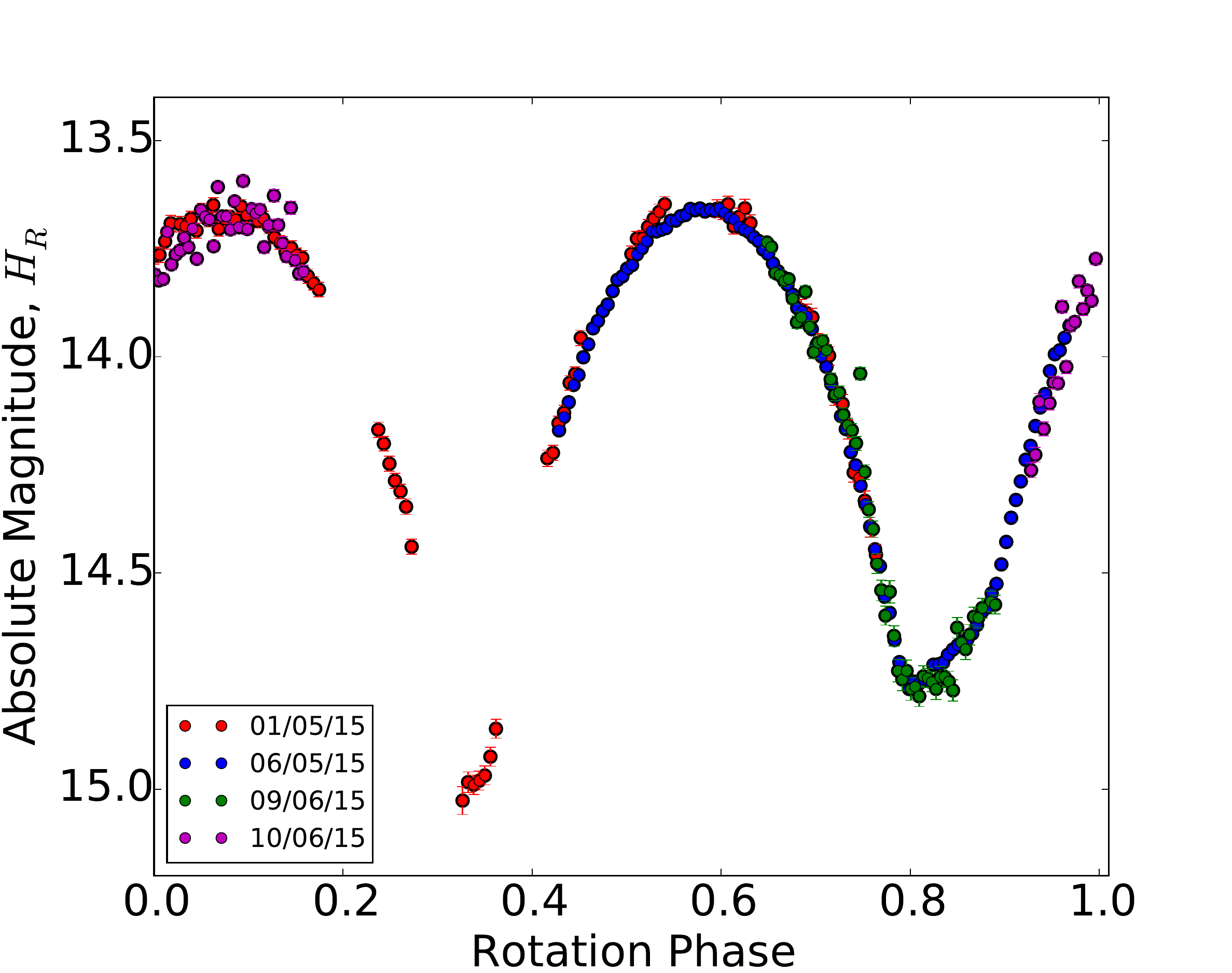}
   \caption{A light curve constructed from photometric data from observations of asteroid 18018 using the Isaac Newton Telescope, University of Hawaii 2.2m Telescope and the William Herschel Telescope in May 2015 and phased to the determined rotation period $P_{r}=5.608$ h.}
\label{fig:18018}
  \end{center}
 \end{figure}

Figure~\ref{fig:18018_ls} shows a best fit rotation period of $P_{r}=5.608\pm 0.002$ h. Light curves folded to this period from the data are given in Figures~\ref{fig:18018}

The light curve for this object shows an apparent amplitude of $A_{obs}=1.33\pm 0.03$ magnitudes, which when corrected for phase angle is $A = 1.00 \pm 0.03$ magnitudes.  Since our observations the Palomar Transient Factory (PTF; \citealt{palomar}) reported a rotational period of $5.6077\pm 0.0015$ h for this object, a value in agreement to within uncertainties our own. From the PTF measurements an amplitude of 0.77 magnitudes was obtained, a smaller change than seen in our observations. If indeed this is a reliable value for the amplitude of the light curve we can infer that the spin pole axis of this object is not aligned to the orbital plane, however, more data would be needed to get a clearer idea of the object's obliquity.



For an object with this rotation rate and amplitude to be explained by a Jacobi ellipsoid we find that the ellipsoid must have axis ratios $1:0.40 \pm 0.01:0.32 \pm 0.01$ and a density of $1700 \pm 50$ \kgm. This density is a plausible value, however, the shape of the object would imply that if this body were strengthless that mass loss should occur. For this object to exist as a single Jacobi ellipsoid some cohesive strength must exist in the object. As for asteroid 45864 we look at the equilibrium figures for ellipsoids with cohesive strength from \cite{holsapple2001}. Using this shape solution and our measured spin period we set a limit on the angle of friction encompassing the values for the majority of asteroids $\phi_{F}<15^{\circ}$. Using these values we find a density range for these equilibrium figures $1200<\rho<4800$ \kgm. Using our Drucker-Prager numerical model across this range we find cohesive strength values in the range $0-160$ Pa.


The Roche model of \cite{lacerda2007} gave a lower limit density for this system of $\rho=3300\pm 100$ \kgm. This density is only believable in the case of a solid chondritic object, not in the case of a rubble pile asteroid. Solid objects like this would not deform due to mutual gravity and hence would not take the form of a Roche tidally deformed bilobed object. As stated in Section~\ref{sub:elong} a binary formed of solid objects without any Roche deformation would be likely to consist of more spherical components and would be unlikely to be able to produce a light curve amplitude as high as the one observed from this object.

Calculating the density of equilibrium figures in the bilobed sequence rotating with this object's rotation period we find a range of solutions $2200<\rho<4400$ \kgm along the sequence. The lower density values are plausible for this asteroid suggesting that a bilobed shape may explain the brightness variation of this object as long as $c'$ does not approach zero. These shapes are also capable of producing light curves matching the observed amplitude of 18018. Thus we consider a bilobed shape to be a possible explanation for the light curve of this object.

In summary, we find possible solutions for both a triaxial ellipsoid with minimum cohesive strength $0-160$ Pa, axis ratio $1:0.40 \pm 0.01:0.32 \pm 0.01$ and density $1200<\rho<4800$ \kgm and for a bilobed object where $c'$ does not approach zero. There is no evidence to suggest either solution to be more likely than the other and thus we consider the simpler case of the Jacobi ellipsoid to be the more viable fit until more data can be obtained.

\subsection{18280 (4245 T-3)}
\label{sub:haa_18280}



\begin{figure}
  \begin{center}
\includegraphics[width=0.5\textwidth]{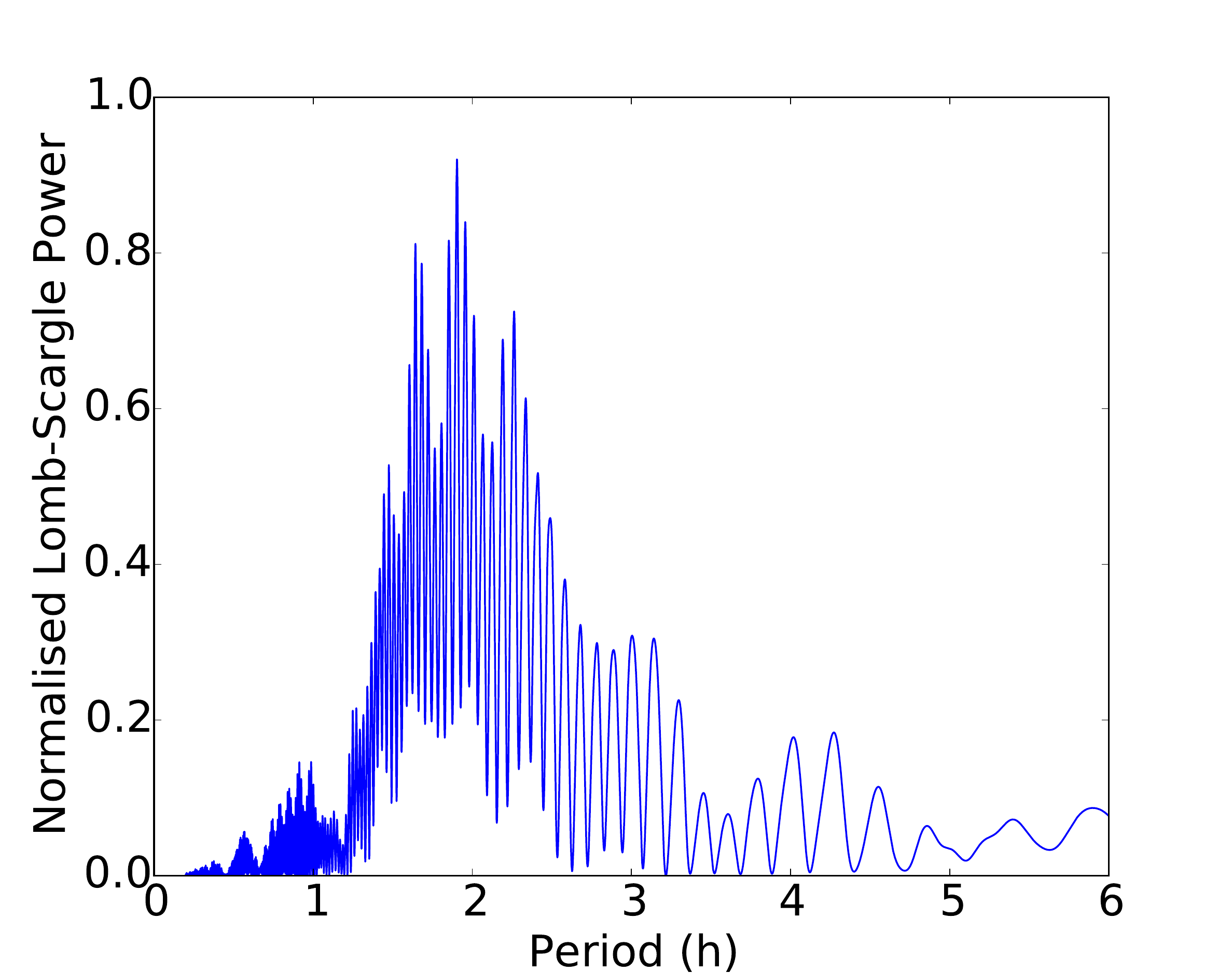}
   \caption{The Lomb-Scargle periodogram generated from the light curve data of 18280}
\label{fig:18280_ls}
  \end{center}
 \end{figure}

The periodogram for this data is shown in Figure~\ref{fig:18280_ls}.


\begin{figure}
  \begin{center}
\includegraphics[width=0.5\textwidth]{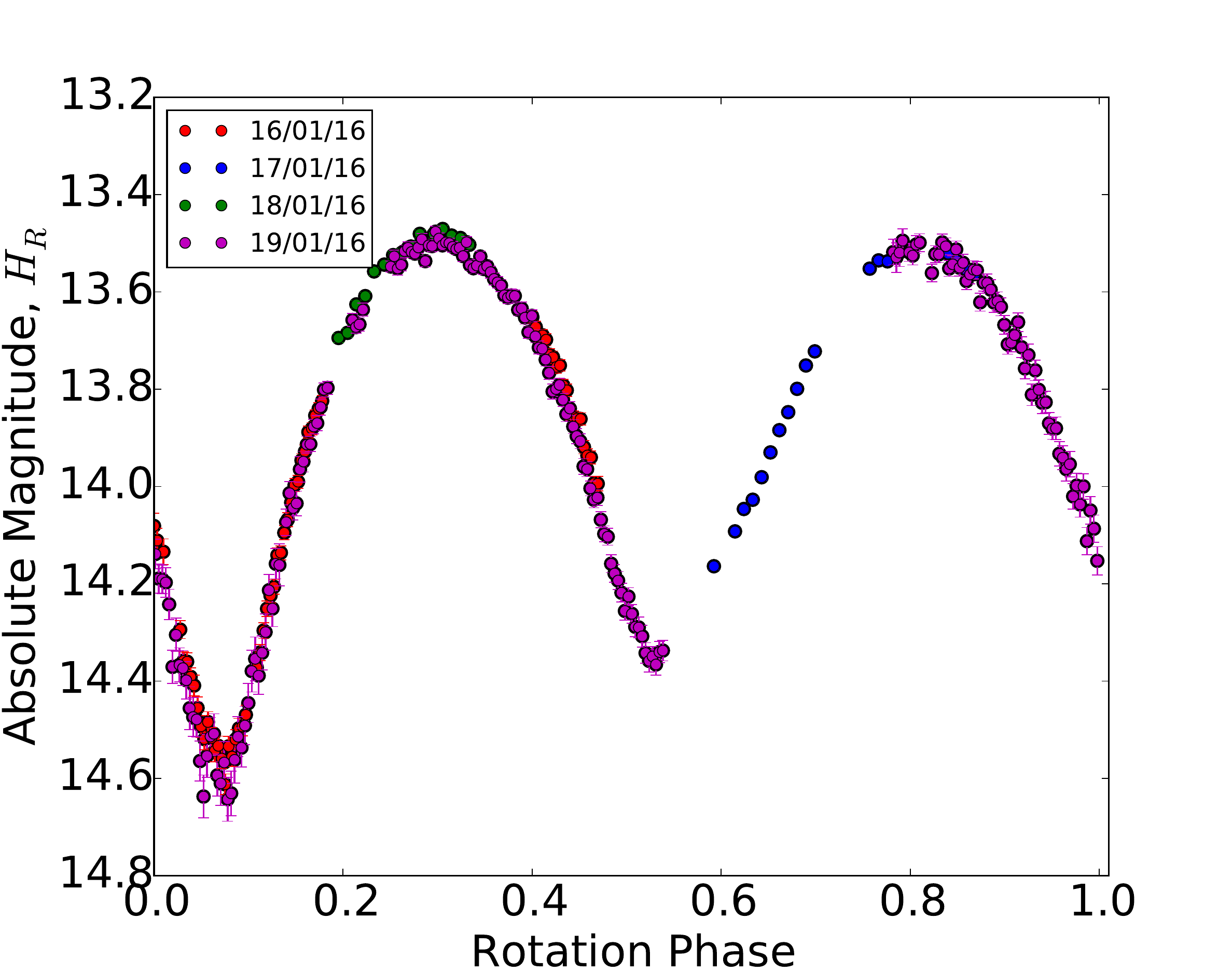}
   \caption{A light curve constructed from photometric data from observations of asteroid 18280 using the Isaac Newton Telescope and the UH2.2m in January 2016 and phased to the determined rotation period $P_{r}=3.808$ h.}
\label{fig:18280}
  \end{center}
 \end{figure}

The best fit period for 18280 was found to be $P_{LS}=1.904 \pm 0.004$ h corresponding to a rotation period of $P_{r}=3.808 \pm 0.008$ h with an observed amplitude of $1.15\pm 0.05$ magnitudes. The corrected amplitude for this object was $0.87 \pm 0.05$ magnitudes, below our cut-off for high amplitude objects. The phased light curve folded to this rotation period is shown in Figure~\ref{fig:18280}

No other light curve data exists for this object at the time this work was written. Although this object does not have an amplitude above our cut-off nor a rotation period shorter than the spin barrier, its above and below average values respectively give us some reason to determine its required strength, if any.

 
 Assuming this object to be a Jacobi ellipsoid we find a best solution for an object with axis ratio of $1:0.45 \pm 0.02:0.35 \pm 0.01$ and a density of $\rho=3500 \pm 100$ \kgm. This shape is slightly less elongated than the limit at which a strengthless body would be expected to undergo mass loss and so this may still be a possible shape for the object. Assuming an angle of friction $\phi_{F}=15^{\circ}$ we find solutions in the density range of $2300<\rho<7900$ \kgm with the upper limit imposed corresponding to the grain density of solid iron, $\rho = 7900$ \kgm. In this range we find required density dependent cohesive strengths in the range $0-400$ Pa, with strengthless solutions for densities in the range $2500 <\rho< 5000$ \kgm.





Therefore we conclude that this object is best explained by a single triaxial ellipsoid with density in the range $2600<\rho<7900$ \kgm ~potentially with some internal cohesive strength if it has a particularly high density, $\rho > 5000$ \kgm.

\subsection{206167 (2002 TS242)}
\label{sub:haa_206167}



\begin{figure}
  \begin{center}
\includegraphics[width=0.5\textwidth]{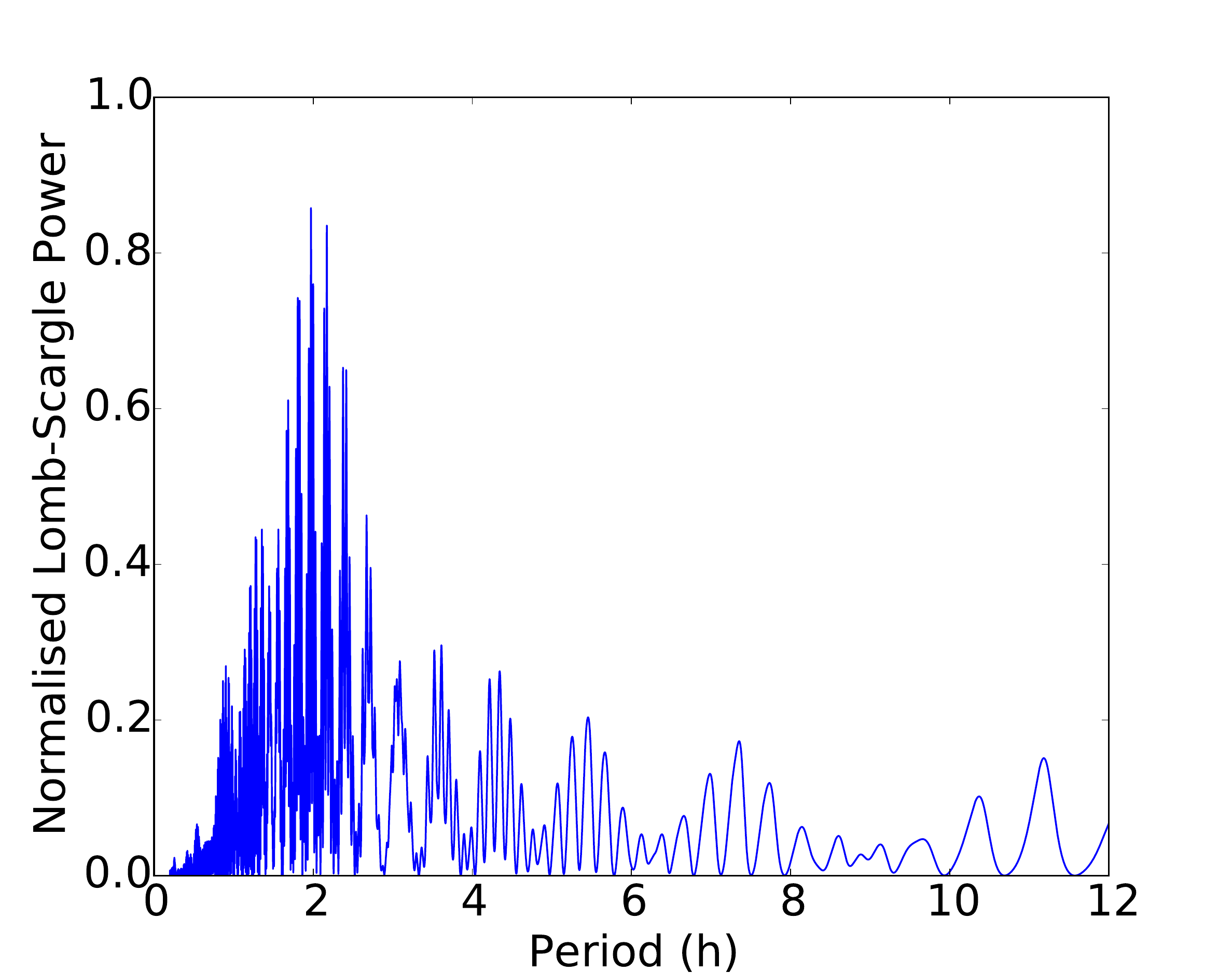}
   \caption{The Lomb-Scargle periodogram generated from the light curve data of 206167 }
\label{fig:206167_ls}
  \end{center}
 \end{figure}

The LS periodogram measured for this asteroid is shown in Figure~\ref{fig:206167_ls}.


\begin{figure}
  \begin{center}
\includegraphics[width=0.5\textwidth]{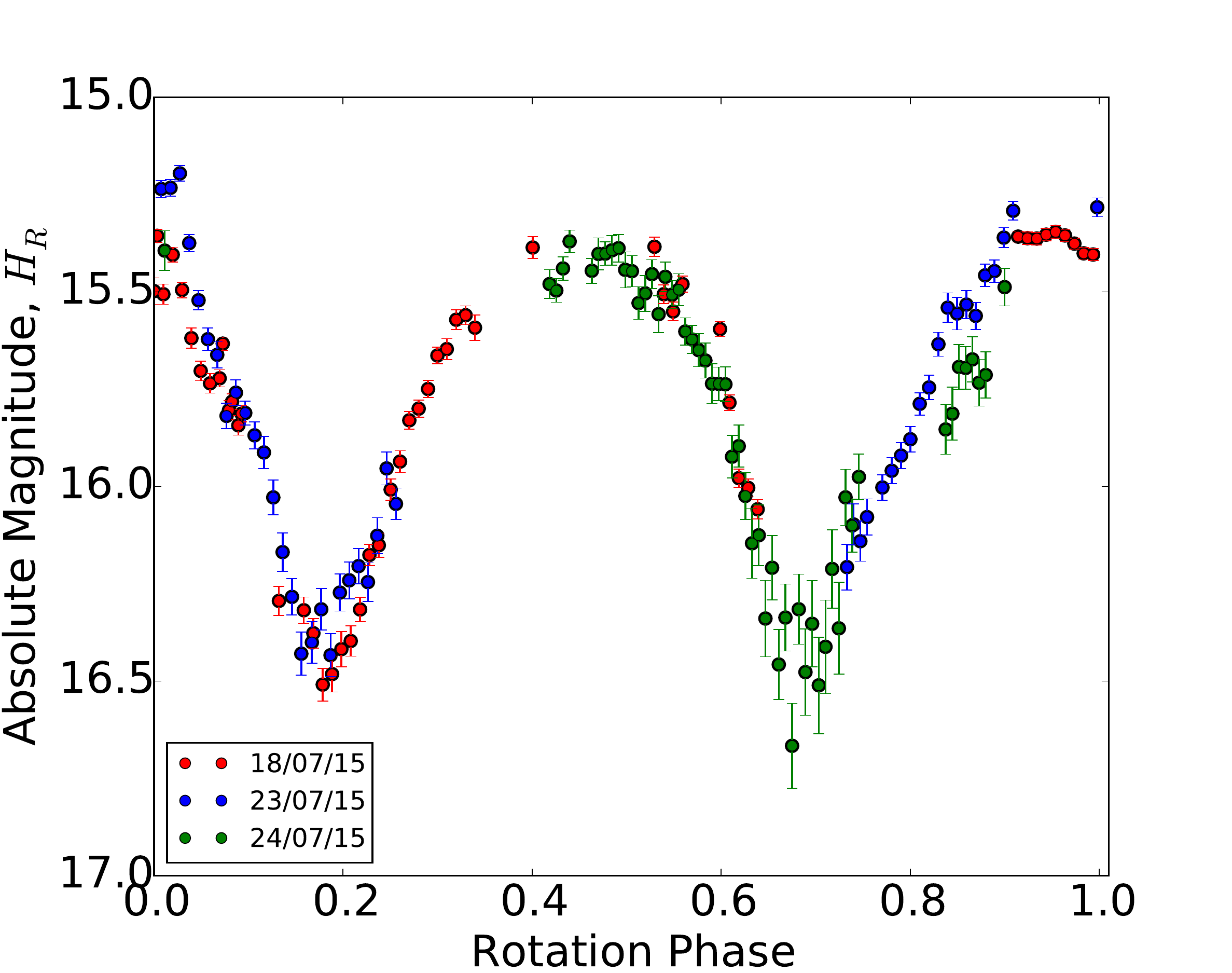}
   \caption{A light curve constructed from photometric data from observations of asteroid 206167 using the NTT and UH2.2m in July 2015 phased to the determined rotation period $P_{r}=3.946$ h.}
\label{fig:206167}
  \end{center}
 \end{figure}

Figure~\ref{fig:206167_ls} shows a best fit rotation period of $P_{r}=3.946\pm 0.006$ h. The light curve of 206167 folded to this rotation period is shown in Figure~\ref{fig:206167}. Considering the central maximum in Figure 13 to be more reliable than the more scattered maximum, we estimate an amplitude of $1.05\pm 0.15$ magnitudes with this value reducing to $0.86\pm 0.12$ magnitudes when phase angle effects are taken into consideration. 

 If we assume the object to be a Jacobi ellipsoid then from \cite{chandrasekhar1969} we find this is best fit by an object with axis ratio $1:0.45 \pm 0.05:0.36 \pm 0.03$. The density of this object was calculated as $\rho=3200 \pm 200$ \kgm. This shape is slightly below the limit at which mass loss would be expected for a strengthless body and so this may still be a possible shape for the object assuming zero cohesive strength. The density obtained however, would imply that this object would have to be a monolith of solid chondrite and would hence not take the form of a Jacobi ellipsoid. We therefore consider it highly unlikely that this object is a strengthless Jacobi ellipsoid. Instead we assume an angle of friction $\phi_{F}=15^{\circ}$ and find a range of possible densities explaining this shape and spin rate $2100-7900$ \kgm. For this density range we then find that cohesive strengths of $0-350$ Pa provide the best solutions for cohesive strength, with strengthless solutions found in the range $2300 < \rho < 3300$ \kgm. 





In conclusion, this asteroid's light curve was best explained by a triaxial ellipsoid with density in the lower end of the range $2600<\rho<7900$ \kgm with axis ratio $1:0.45 \pm 0.05:0.35 \pm 0.03$ and density-dependent required cohesive strength $0-350$ Pa. 

\subsection{15613 (2000 GH136)}
\label{sub:haa_15613}



\begin{figure}
  \begin{center}
\includegraphics[width=0.5\textwidth]{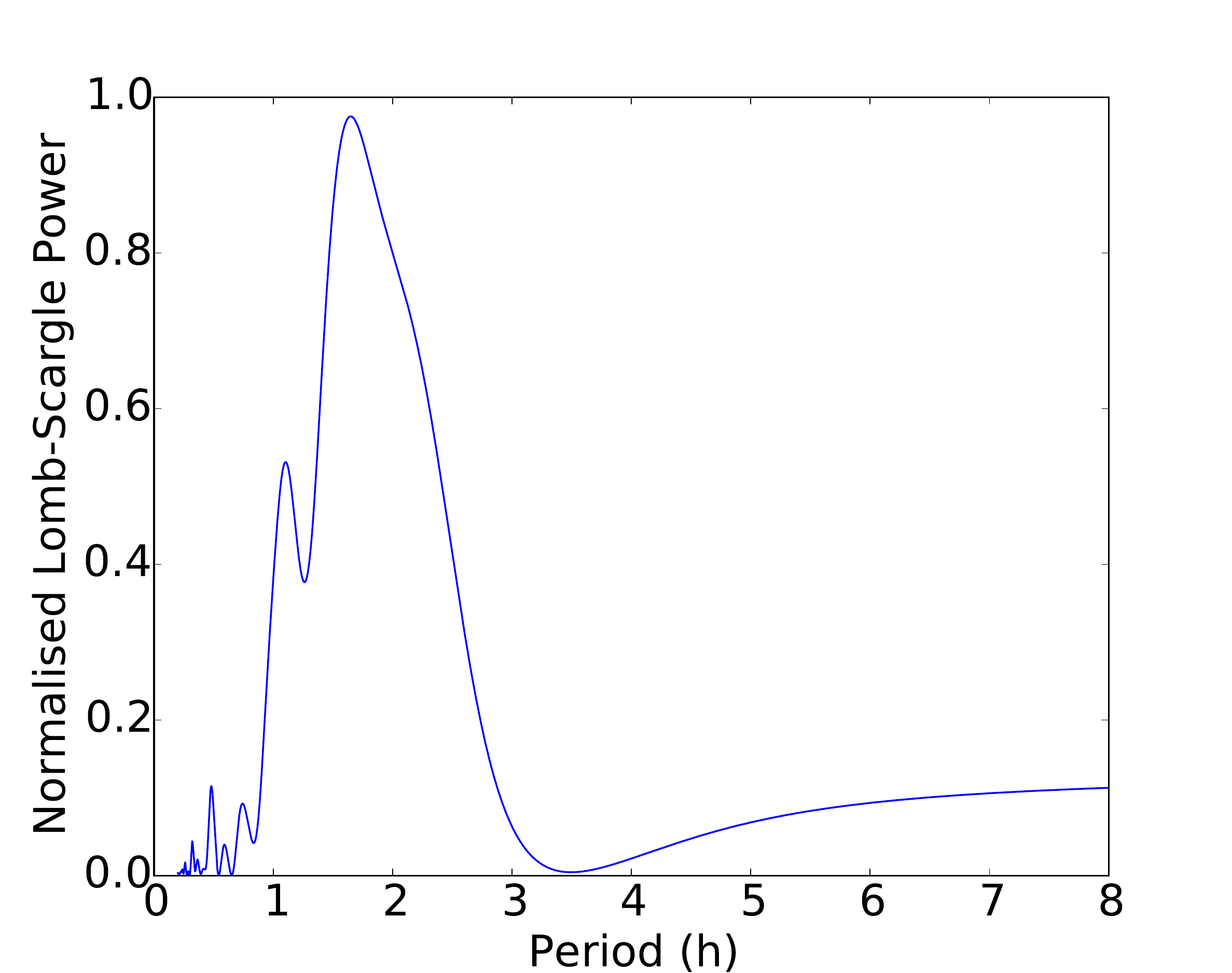}
   \caption{The Lomb-Scargle periodogram generated from the light curve data of 15613}
\label{fig:15613_ls}
  \end{center}
 \end{figure}

This object was observed for only one night late in the project on 19/04/16 using the INT during which $47$ $60$-second exposures were taken. The LS periodogram for the data taken is given as Figure~\ref{fig:15613_ls}.

\begin{figure}
  \begin{center}
\includegraphics[width=0.5\textwidth]{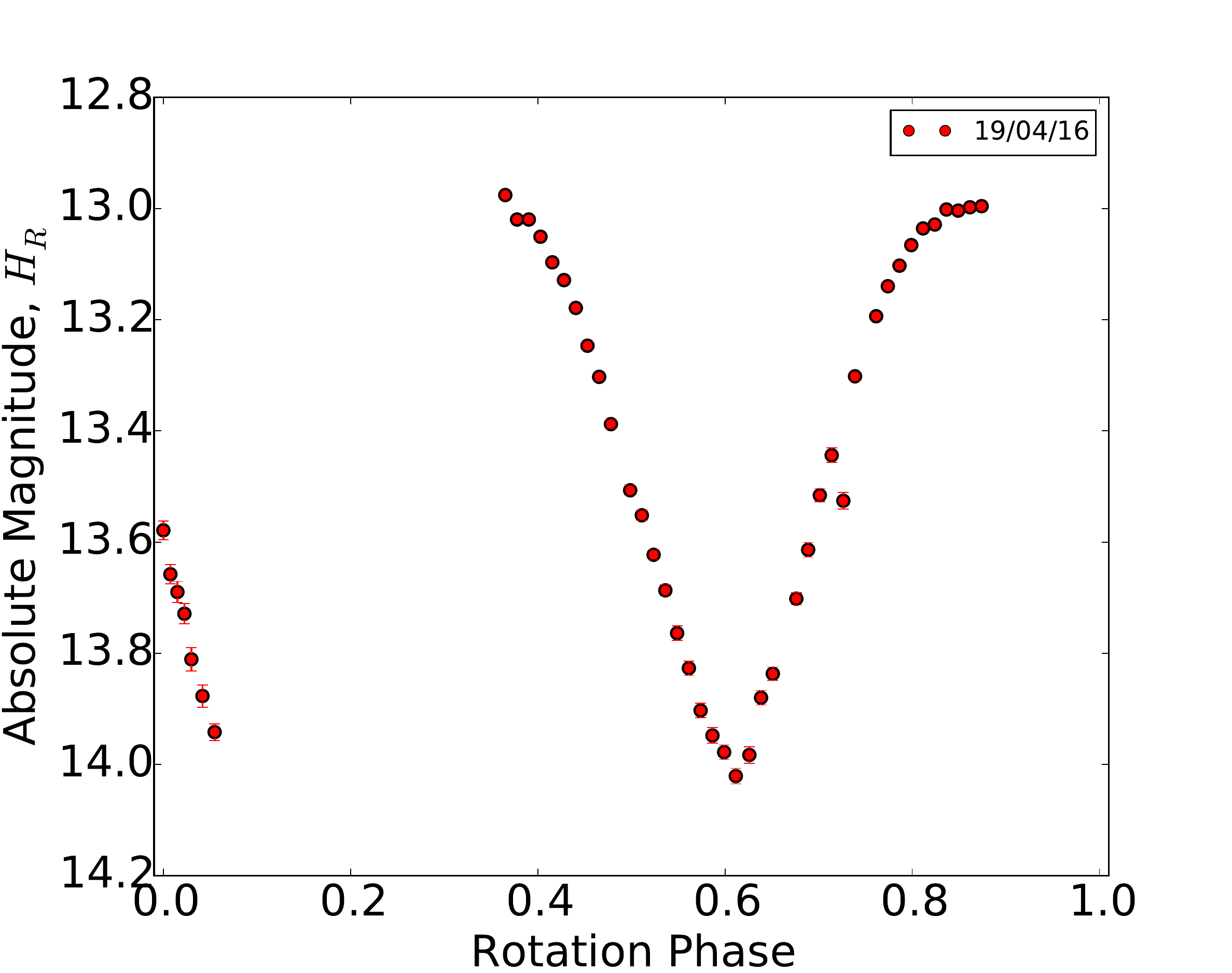}
   \caption{A partial light curve constructed from photometric data from observations of asteroid 15613 using the Isaac Newton Telescope in April 2016 phased to the determined rotation period $P_{r}=3.3$ h.}
\label{fig:15613}
  \end{center}
 \end{figure}

The best fit rotation period from Figure~\ref{fig:15613_ls} was $P_{r}=3.3 \pm 0.4$ h. If this period determination is accurate then the data we have obtained for this object does not represent a full light curve and further observations will be needed to confirm an accurate period value. The partial light curve obtained folded to the best fit period is given in Figure~\ref{fig:15613}.

Even in this partial light curve an apparent amplitude of $A=1.02\pm 0.02$ magnitude is seen, although we would require a full light curve containing the second light curve minimum before deeming this the real amplitude. This lower amplitude limit becomes $0.78 \pm 0.02$ mag when the phase-amplitude relationship is accounted for. The Palomar Transient Factory (\citealt{palomar}) identified this object as having a rotation period of $3.4184 \pm 0.0004$ h, in agreement with the value we obtain from the LS periodogram. As in the case of 18280, although this object does not meet our cut-off for high amplitude objects we calculate its required cohesive strength due to its relatively short rotation period.



In summary we find that the best solution for a Jacobi ellipsoid producing this light curve is given by the axis ratio $1:0.49 \pm 0.01:0.38 \pm 0.01$ with a density $\rho=4600 \pm 1100$ \kgm. Assuming an angle of friction $\phi_{F}=15^{\circ}$ we find solutions in the density range of $3200<\rho<7900$ \kgm with the upper limit imposed corresponding to the grain density of solid iron. In this range we find that the body can be strengthless if its density is $\rho > 3700$ \kgm, for lower densities than this strength of up to 250 Pa may be required. It is worth noting that this analysis is based on a partial light curve and for a more reliable conclusion, further observations of this object would be required.

\subsection{49257 (1998 TJ31)}
\label{sub:haa_49257}



\begin{figure}
  \begin{center}
\includegraphics[width=0.5\textwidth]{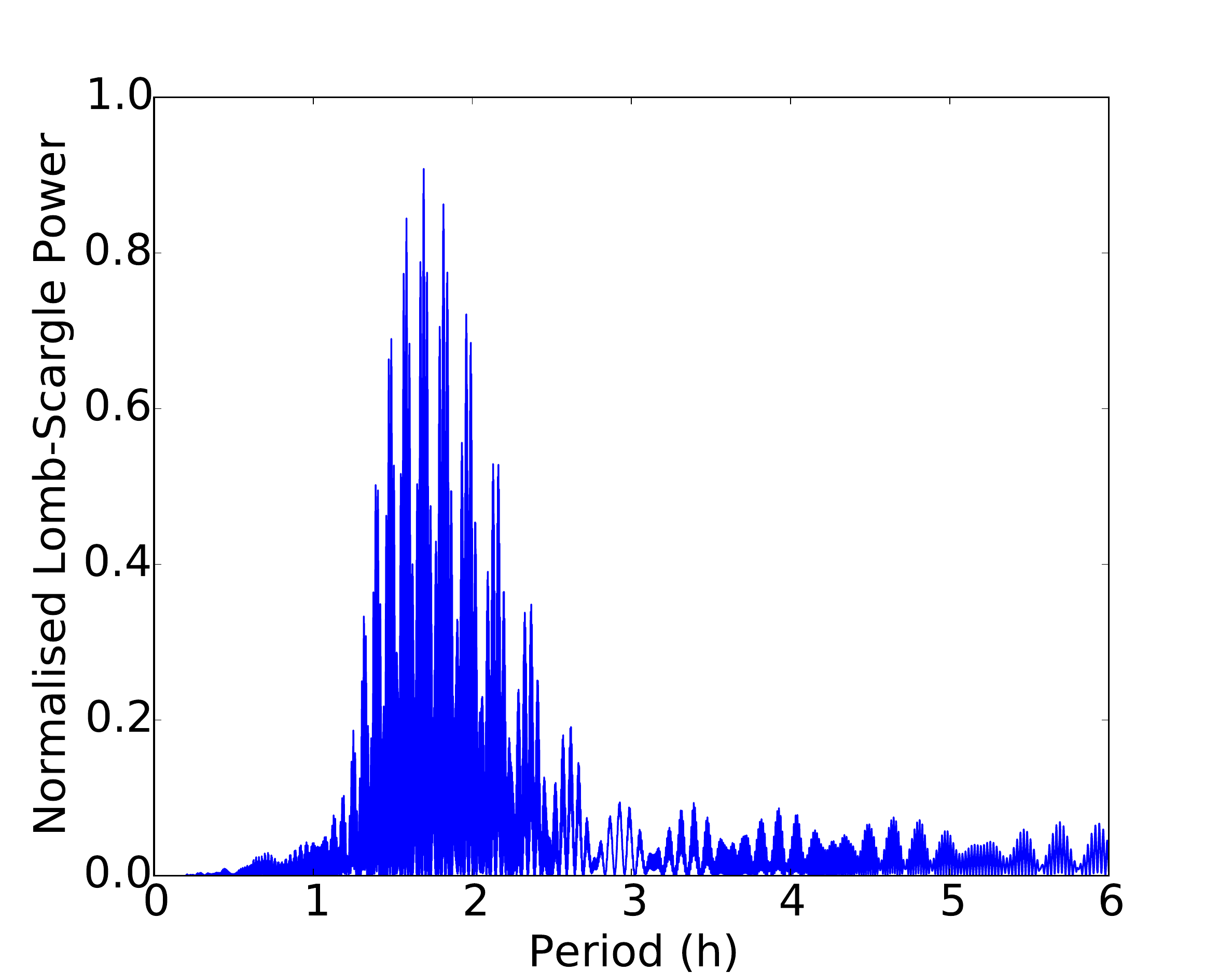}
   \caption{The Lomb-Scargle periodogram generated from the light curve data of 49257}
\label{fig:49257_ls}
  \end{center}
 \end{figure}

The Lomb-Scargle periodogram for this object is given in Figure~\ref{fig:49257_ls}.


Figure~\ref{fig:49257_ls} shows a best fit rotation period of $P_{r}=3.3896 \pm 0.0002$ h. The light curves for this object phased to this period are given in Figures~\ref{fig:49257_lc1} and \ref{fig:49257_lc2}.

\begin{figure}
  \begin{center}
\includegraphics[width=0.5\textwidth]{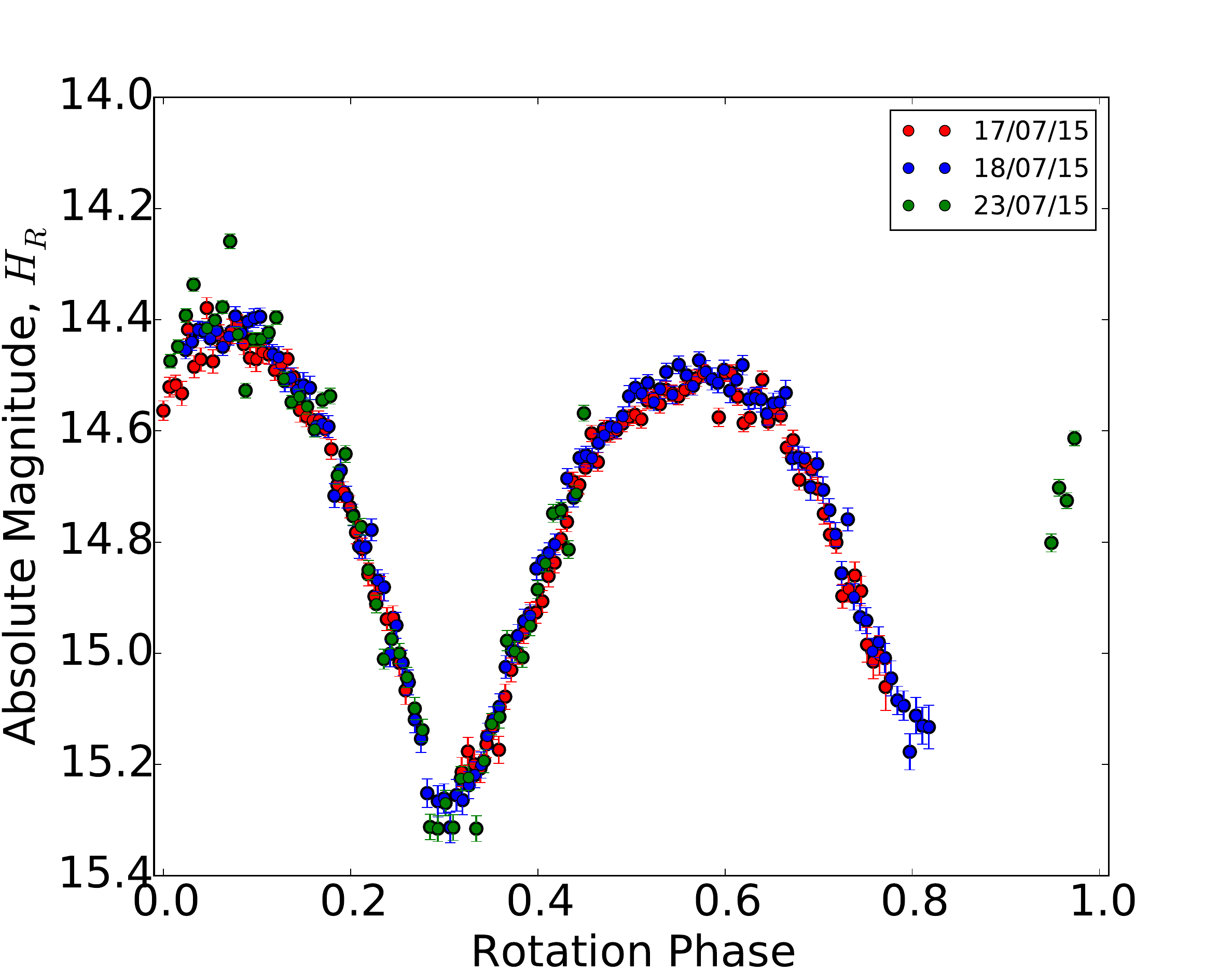}
   \caption{A light curve constructed from photometric data from observations of asteroid 49257 using the NTT and UH2.2m in July 2015 phase folded to the determined rotation period $P_{r}=3.3896$ h. }
\label{fig:49257_lc1}
  \end{center}
 \end{figure}

\begin{figure}
  \begin{center}
\includegraphics[width=0.5\textwidth]{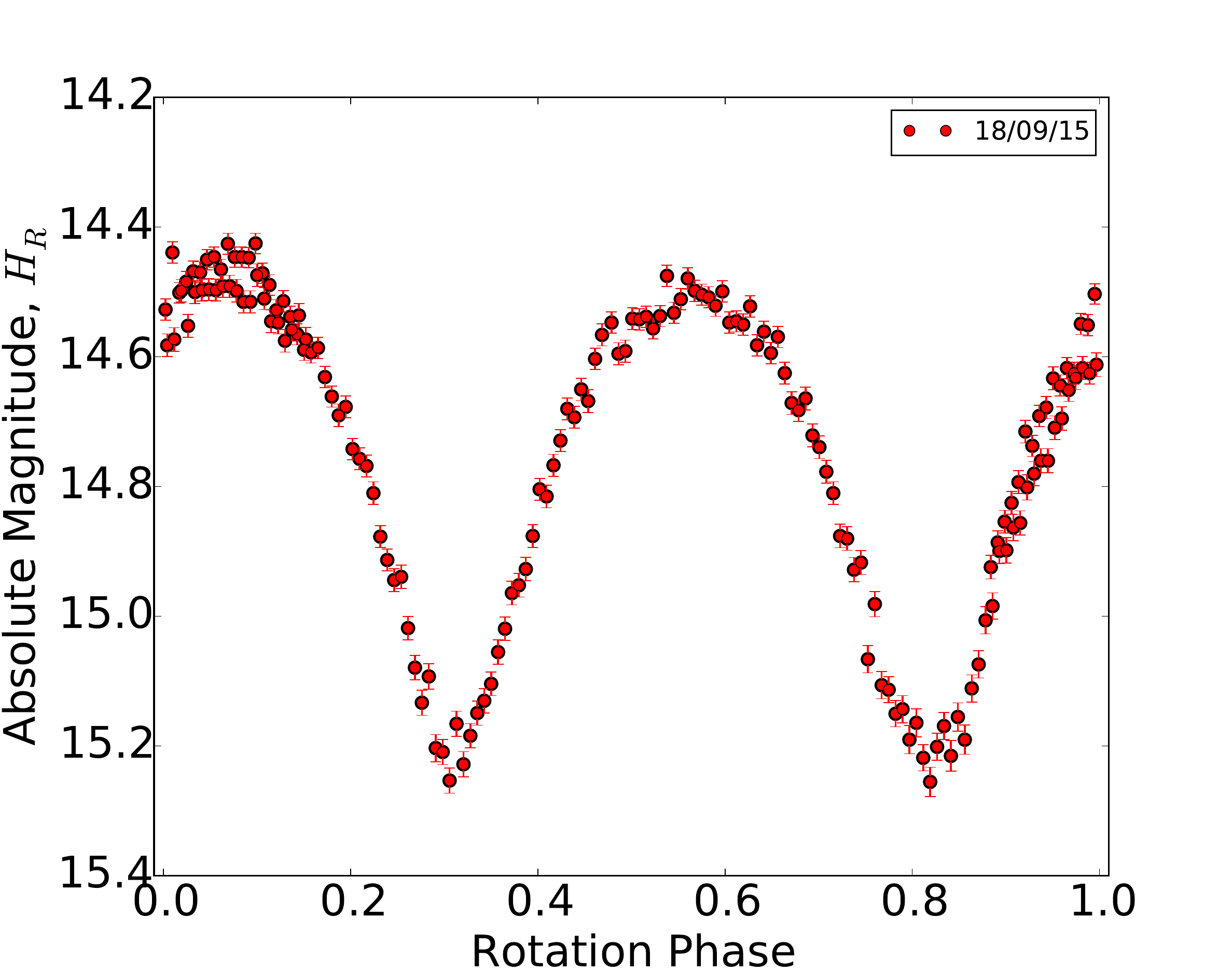}
   \caption{A light curve constructed from photometric data from observations of asteroid 49257 using the UH2.2m in September 2015. }
\label{fig:49257_lc2}
  \end{center}
 \end{figure}

\begin{figure}
  \begin{center}
\includegraphics[width=0.5\textwidth]{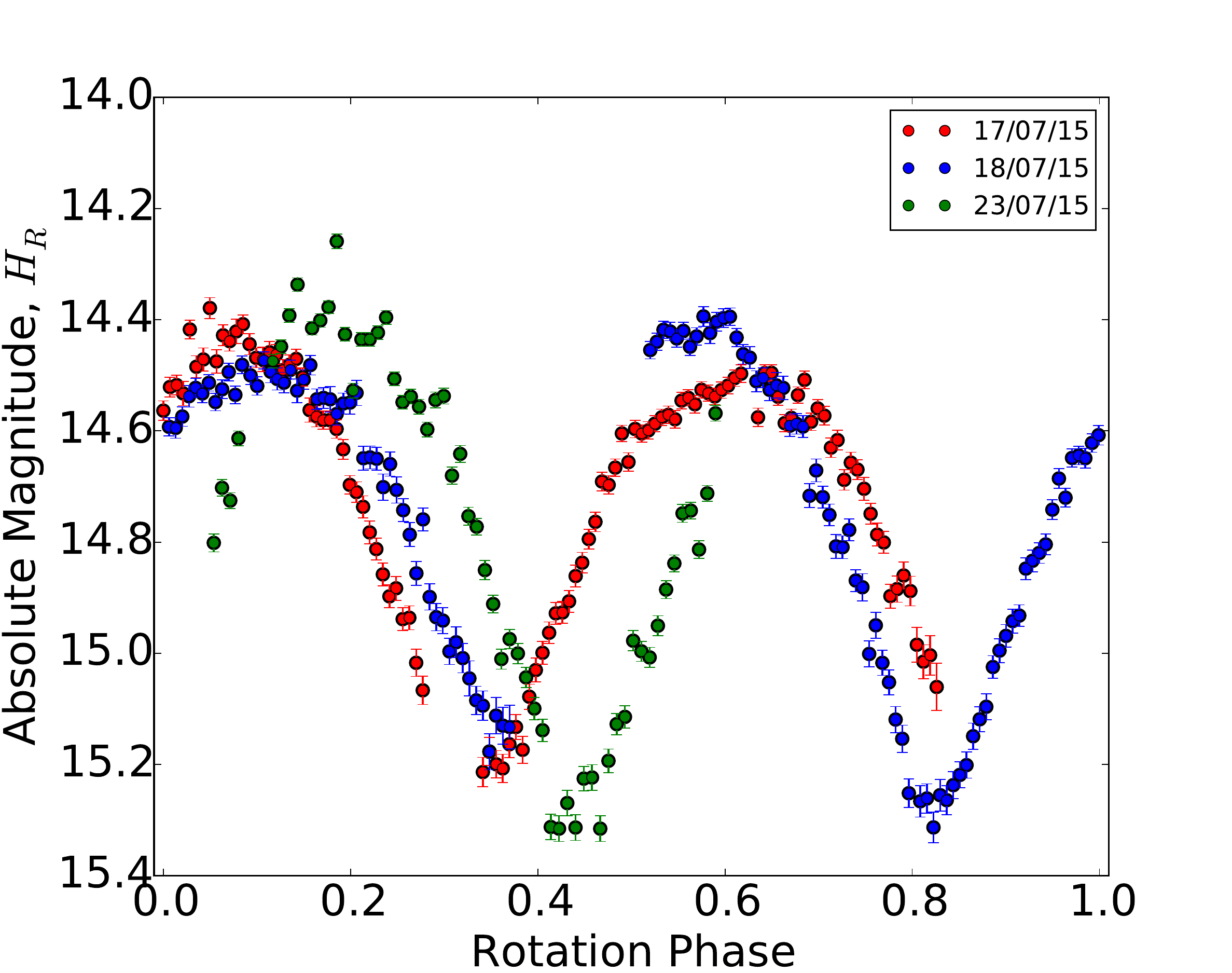}
   \caption{A light curve constructed from photometric data from observations of asteroid 49257 using the NTT and UH2.2m in July 2015 phase folded to the rotation period determined by the Palomar Transient Factory, $P_{PTF}=3.1661$ h. }
\label{fig:49257_ptf}
  \end{center}
 \end{figure}

The Palomar Transient Factory (\citealt{palomar}) reported a rotation period for this object of $3.1661 \pm 0.0005$ h which differs from our obtained value by $7\%$. This period is listed by the LCDB as having a quality code of 2 defined as a "result based on less than full coverage. Period may be wrong by 30 percent or ambiguous" (\citealt{warner2009}). For comparison we include a phased light curve of our July 2015 photometric data folded to the rotation period found by PTF in Figure~\ref{fig:49257_ptf}. The phase corrected amplitude of this object was $0.58\pm 0.05$ magnitudes, below the cut-off criteria for magnitude above which we consider an asteroid to be truly high-amplitude. Despite this, we serendipitously obtained further observations of this object at different orbital geometry which allowed us to obtain shape and spin data using lightcurve inversion modelling and found a best fit model showing that this object is expected to have a higher maximum amplitude than our cut-off. The details of this process will be covered in Section~\ref{sec:shape}.


From our data we see that the period determined by the Palomar Transient Factory does not appear to fit our data, and thus we reject it in favour of our own value $P_{rot}=3.3896 \pm 0.0002$ h.

\subsection{Remaining objects observed}
\label{subsec:haa_remain}

In addition to the objects considered in the previous section, we observed a further 14 targets throughout the course of the project. The orbital parameters as well as the determined periods and amplitudes are given in Table~\ref{table:remain}. Of these objects, only 2 can truly be considered to have been false positives in our target list as their rotation period determined either by us or by PTF is too long to have ever produced a change of 0.3 magnitudes in 15 minutes. For the remainder of our targets it is impossible to state definitively whether these targets are truly high amplitude or not as we don't know anything about their spin pole orientation. In these cases their rotational periods are sufficiently short to have produced a large variation in magnitude if these objects have a higher amplitude than measured from our observations. This could be as a result of PS1 observing the target equator-on or at least closer to this angle than for our own observations.

\begin{table}
\caption{Overview of other objects observed during this project.} 
\centering 
\begin{tabular}{c c c c} 
\hline\hline 
Object & Period (h) & Amplitude (mag) \\  
\hline 
23314 & $6.2104 \pm 0.0004$ & 0.87\\
81847 & $4.930 \pm 0.002$ & 0.86\\
140666	& $3.5 \pm 0.1$	& 0.7 \\
55125	& $5 \pm 0.2$	& 0.7  \\
5929	& $3.76 \pm 0.01$	& 0.6  \\
3764	& $>4$	& $>$0.5  \\
12582	& $\sim 4$	& 0.4  \\
172388	& $\sim8-10$ & 0.3 \\
56656	& $2.7$	& 0.25 \\
114372	& $~3.5$	& 0.2  \\
112815	& -	& -	\\
173046	& -	& -	\\
15354	& -	& -	\\
85298	& -	& -	
\\\hline 
\end{tabular}
\label{table:remain} 
\end{table}


Although we may find an object to have an apparent amplitude of less than 1 mag, the object may still be high amplitude and the observed amplitude may be due to orbital geometry. As such we are unable to definitively rule out objects as high-amplitude without further data. In the case of all targets observed we require a second set of observations at different orbital geometries ideally $90^{\circ}$ along their orbit since previously observed in order to show that they are not high-amplitude asteroids. Using the shape distribution for small main belt asteroids from \cite{mcneill2016} we estimate that as many as $0.6\%$ of objects in this size range may produce light curve amplitudes $A \geq 1.0$ mag.


\subsection{Limitations}
\label{subsec:haa_limit}

Our selection criteria and observational strategy are not without their limitations. As stated earlier, from the sparse data alone it is impossible to determine in advance whether an object showing large rates of change in magnitude is an object with a fast rotation rate or a high amplitude light curve. Knowing this would allow us to much more easily optimise our observing strategy for each object and make more efficient use of our time. However, with the data available this is an unsolvable problem so we have simply had to adapt 'on the fly' during our observing runs.

It is also not immediately clear when we observe an object that we can definitively rule out that it is a high amplitude object. For some of our remaining objects we obtained light curve amplitudes $A<0.8$ magnitudes, but without further data there is no way to tell whether these objects actually have a lower amplitude than we are interested in, or if the observing geometry is simply such that we are not seeing the full extent of the amplitude in our light curve due to the spin pole axis of the asteroid. Following up on these objects would be time consuming and potentially inefficient. We also looked at the difference in true anomaly between the objects when they were observed by us at the telescopes and when they were observed by Pan-STARRS. If there is a change approaching $90^{\circ}$ or $270^{\circ}$ then this may suggest that the difference in observed amplitude is due to orbital geometry but there is still no way to conclusively state this without further data. During the project to date we did not perform any follow up observations for objects with observed $A<0.8$ mag but this may present a useful avenue for further work as similar to the case of 5929 we would expect that some of these seemingly less interesting objects are in fact high amplitude after all. It is worth noting that any objects for which we determined a period $P>10$ h is unlikely to really be a high amplitude object if we see only a low amplitude light curve. This is because for an object with a period greater than this to produce the sort of magnitude variations we look for as our selection criteria an unphysical amplitude would be required. It is not known whether these large changes are simply the effect of contamination or are a real effect due to some outside factor, but for our purposes we consider these objects to be false positives.

An interesting finding also is that there appeared to be no clear correlation between the maximum $\dot{m}$ found from the Pan-STARRS data and the extent of the light curve amplitude observed, at least at higher values. Initially we used this value to determine the order we would proceed through our target list but after 2 observing runs and no clear correlation we altered our strategy to instead try and observe as many of our targets as possible rather than simply focusing on a single target which showed a large value of $\lvert\dot{m}\rvert$ in the survey data. Instead it appears as though the likelihood of an object being high amplitude is independent of the maximum rate of change in magnitude observed where $\dot{m}_{max} > 0.6$ mag per 15 minutes. This is likely due to the difference in orbital geometry between the PS1 detections of the object and our follow up observations.

\subsection{Summary of observations}

\begin{figure}
  \begin{center}
\includegraphics[width=0.5\textwidth]{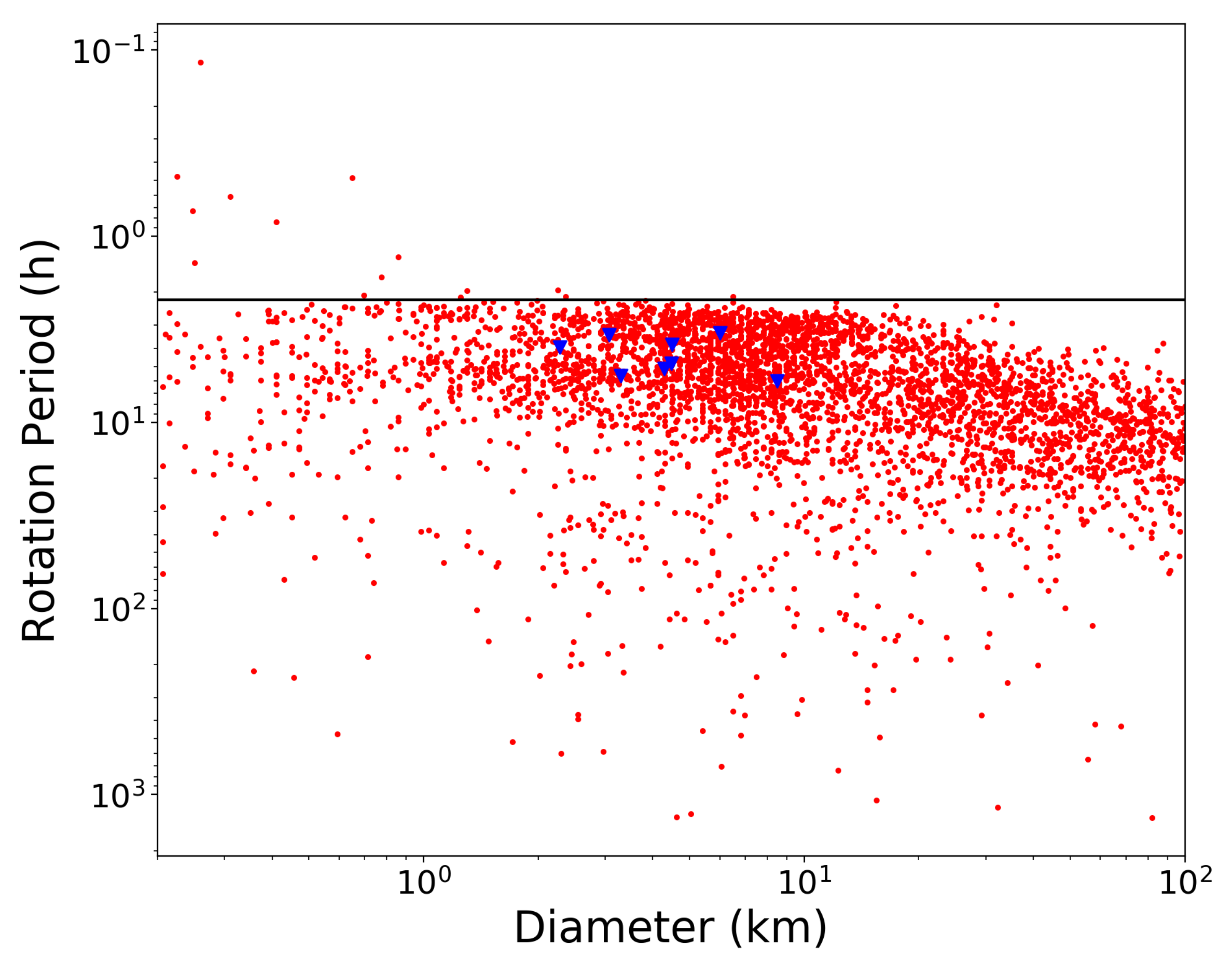}
   \caption{The rotation period and diameter of the 8 primary objects (blue points) analysed in Section 5 shown on the same axes as the corresponding data from the Light Curve Database for all objects $0.2 < D < 100$ km.}
\label{fig:diamperiod2018}
  \end{center}
 \end{figure}

In this project we have carried out observations of 22 asteroids from our target selection list. In this section we have detailed the observations and analysis of 8 objects considered to be closest to our definition of 'extreme asteroids'. Figure~\ref{fig:diamperiod2018} shows these objects plotted on the same period-diameter axes as the information from the Light Curve Database shown in Figure~\ref{fig:lcdb_spin}. Clearly none of our objects have periods shorter than the spin-barrier and hence none can be considered large super-fast rotators.

\begin{figure}
  \begin{center}
\includegraphics[width=0.5\textwidth]{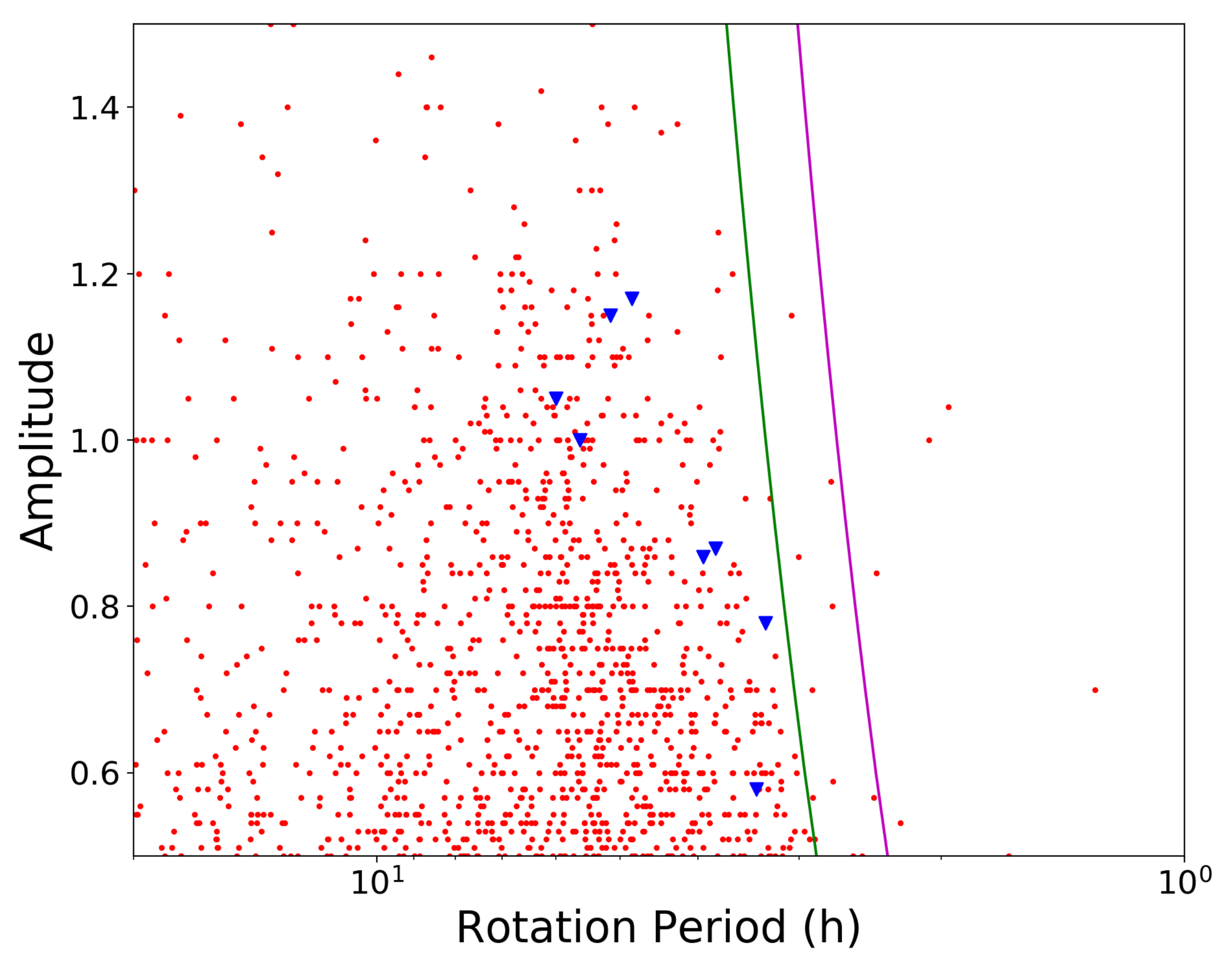}
   \caption{The amplitude and rotation period of the 8 primary objects (blue points) analysed in Section 5 compared to known data from the Light Curve Database for objects $0.2 < D < 100$ km. Note that the amplitudes from the LCDB are not corrected for phase angle. The green and magenta lines show the critical spin period for objects with bulk densities $\rho = 2000$ \kgm ~and $\rho = 3000$ \kgm ~respectively.}
\label{fig:diamamp2018}
  \end{center}
 \end{figure}

Figure~\ref{fig:diamamp2018} shows the amplitudes and rotation periods of these 8 objects along with the same information for objects in the Light Curve Database. Also shown are the critical spin periods for objects with $\rho = 2000$ \kgm and $\rho = 3000$ \kgm. Here we see that the 8 objects studied here do not exceed the given critical spin rates given their density. We find that depending on their density that these objects may still require some cohesive strength but these values are similar to or lower than known values for asteroid cohesive strengths.

From the data available to us from the Pan-STARRS 1 Survey we have not identified any objects with super-fast rotation or high amplitude requiring unusually high cohesive strength.

\section{Shape Modelling}
\label{sec:shape}

\subsection{Light Curve Inversion}
\label{sec:inv_model}

Only a small proportion of asteroids have known shapes as obtained by radar observations or resolved from images. These methods are also preferentially biased toward large and/or nearby objects. For most objects the best method of obtaining shape and spin pole information is through the use of light curve inversion. This method allows for an approximate shape to be determined as radar imaging of main belt asteroids has revealed objects with large-scale concave features which can not be reproduced by light curve inversion (\citealt{hudson1999}; \citealt{hudson2000}). In addition to this it is difficult to determine a unique spin-pole solution as two solutions will generally be found with approximately the same latitude and longitude values separated by $180^{\circ}$. A typical approach to light curve inversion is to place shape constraints on the modelled objects, for instance allowing only minor deviations from a triaxial ellipsoid. To obtain shape and spin models for those asteroids with sufficient observations we utilise the light curve inversion method devised by \cite{kaas2001} using the software package developed by \cite{durech2010}. 

\begin{equation} 
dL=S(\mu,\mu_{0})\bar{\omega} ds
\label{eqn:dl}
\end{equation}

The total brightness of an asteroid, L,  can be expressed as a function of the shape, surface scattering and rotation state of the object. The change in the total brightness of the asteroid over time is dependent on the evolution of each of these parameter sets with time. By summing the model values of $dL$ from each surface facet of the modelled object, the value of the brightness for a series of variable parameters, L$_{mod}$ can be determined. Comparing this model value with the observed light curve and minimising chi-squared for a variety of shape and scattering configurations as $\chi^{2}=||L_{obs} - L_{mod}|| ^{2}$.

During this investigation we obtained enough observations at different orbital longitudes to allow shape modelling for 3 of our targets: 45864, 206167 and 49257. In this section the preliminary shape results for each of these objects are discussed.

\subsection{45864 (2000 UO97)}
\label{subsec:inv_45864}

\begin{figure}
  \begin{center}
\includegraphics[width=0.5\textwidth]{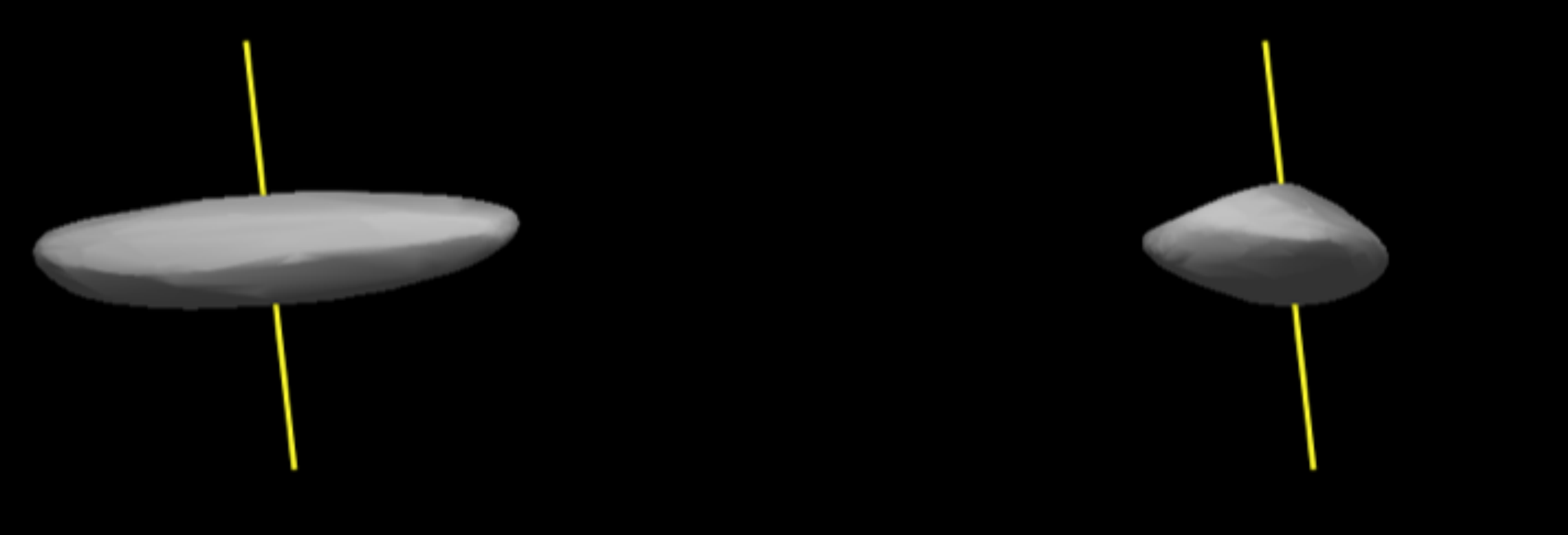}
   \caption{The output shape model of 45864 at two points about its rotation separated by $90^{\circ}$. The object is aligned at its obtained spin pole latitude $\beta=-82^{\circ}$.}
\label{fig:45864_par}
  \end{center}
 \end{figure}

Using light curve inversion the best fit shape generated for 45864 is presented as Figure~\ref{fig:45864_par}. The model object has approximate axis ratios $1:0.31:0.31$ and its best fit spin pole longitude and latitude are given by $\beta= 218\pm 5^{\circ} , \theta=-82\pm 5^{\circ}$. This shape is less elongated in terms of axis ratio than the object determined in Section~\ref{sub:haa_45864} as this inversion method does not assume a triaxial ellipsoid, and additionally accounts for illumination effects such as limb darkening. This less elongated shape would allow the object to resist rotational fission at lower strength values than those obtained for the previous shape. The density range for the object would presumably also change but without a more advanced model it is not possible to calculate these values, as the work in Section~\ref{sub:haa_45864} assumed a simple triaxial ellipsoid.

\begin{figure}
  \begin{center}
\includegraphics[width=0.5\textwidth]{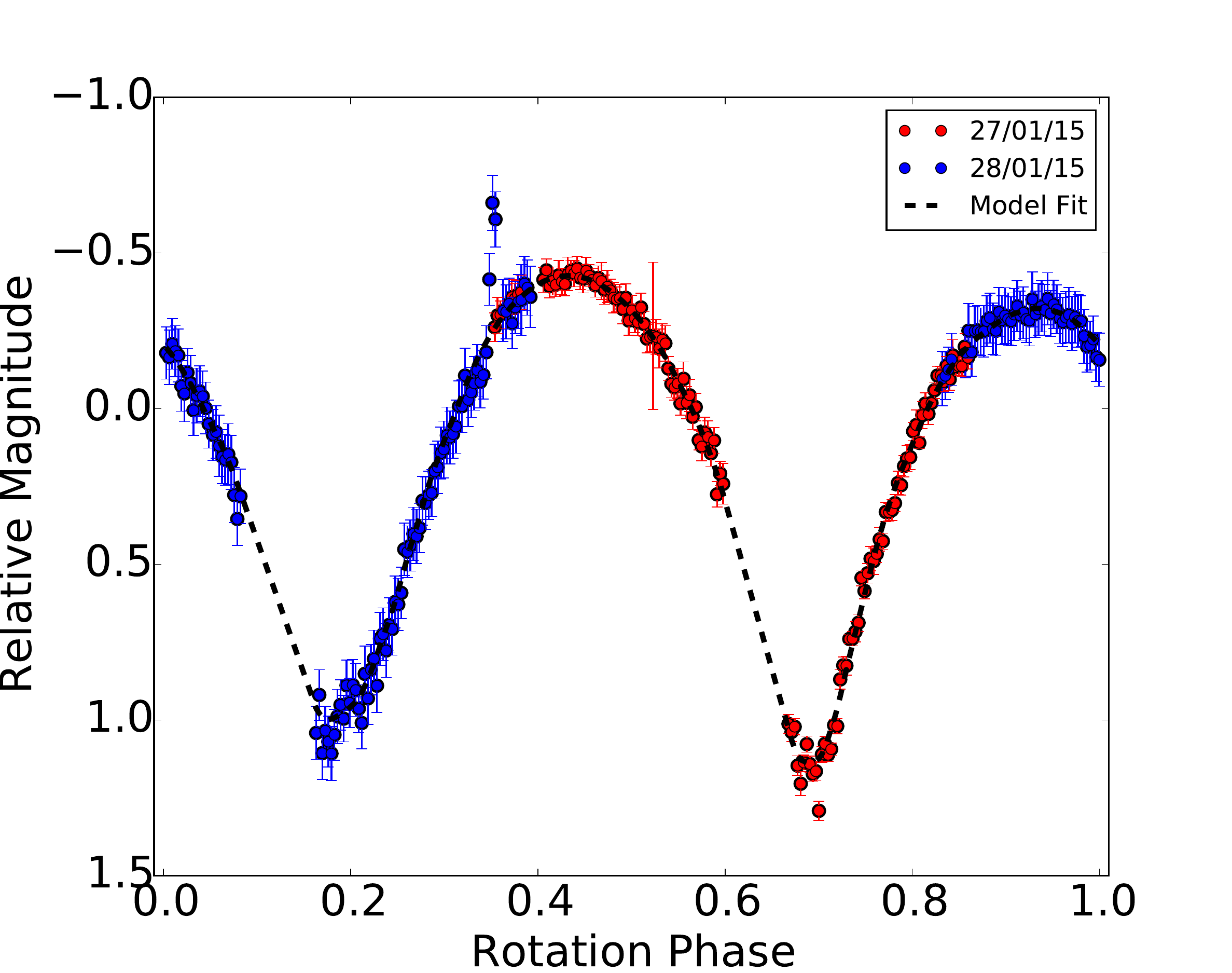}
   \caption{The same light curve as shown in Figure 4.3. Th e black line represents the estimated photometric light curve of the output shape model seen in Figure~\ref{fig:45864_par}}
\label{fig:45864_lco1}
  \end{center}
 \end{figure}

The best fit light curve produced by this model is plotted along with the observed light curves from the January 2015 observations in Figure~\ref{fig:45864_lco1}.

%

\begin{figure}
  \begin{center}
\includegraphics[width=0.5\textwidth]{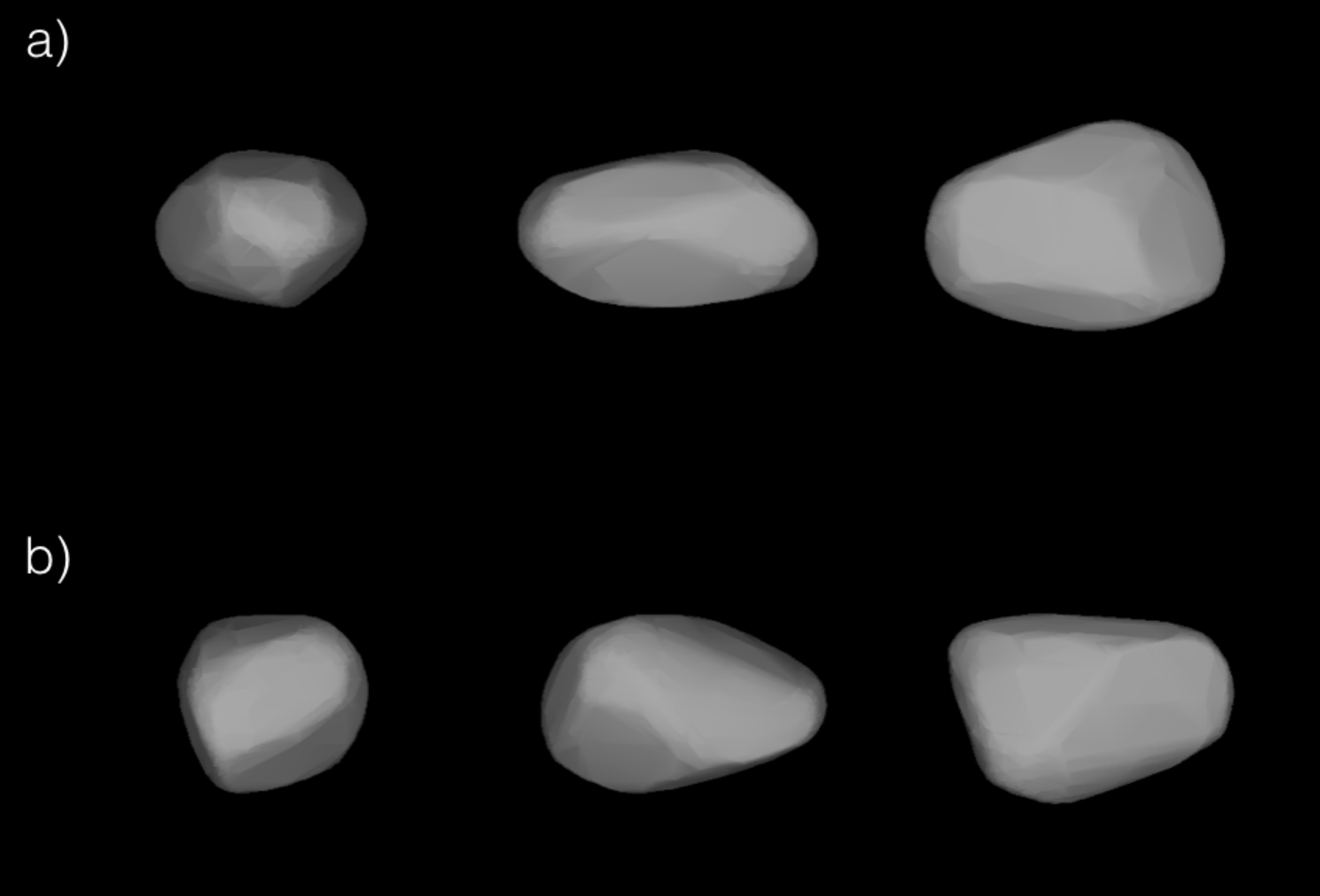}
   \caption{The best fit shape solutions for 45864 as listed in DAMIT (\citealt{durech2016}). a) A best fit shape with spin pole longitude and latitude $\lambda=179^{\circ}, \beta=-84^{\circ}$. b) A best fit shape with spin pole longitude and latitude $\lambda=44^{\circ}, \beta=-66^{\circ}$}
\label{fig:45864_damit}
  \end{center}
 \end{figure}

\cite{durech2016} used sparse photometry from archival data from the Lowell Observatory to construct shape and spin models for 328 new asteroids in addition to the $\sim 350$ which were previously available on the Database of Asteroid Models from Inversion Techniques (DAMIT; \citealt{durech2010}). This added sample included best fit models of 45864 which are shown in Figure~\ref{fig:45864_damit} a) and b). The object was determined to have spin parameters $\lambda=179^{\circ}, \beta=-84^{\circ}$ for a) and $\lambda=44^{\circ}, \beta=-66^{\circ}$ for b). Neither of these shape models would generate the light curves obtained from our observations and given the dense nature of our photometry we favour our own fit. The spin pole latitude obtained from our model, $\beta=-82\pm 5^{\circ}$, is in excellent agreement with the first Durech fit. Further data will allow us to constrain the shape and spin state information of this object even further and we hope to obtain observations at at least one more apparition in the future.

\subsection{206167 (2002 TS242)}
\label{subsec:inv_206167}

\begin{figure}
  \begin{center}
\includegraphics[width=0.5\textwidth]{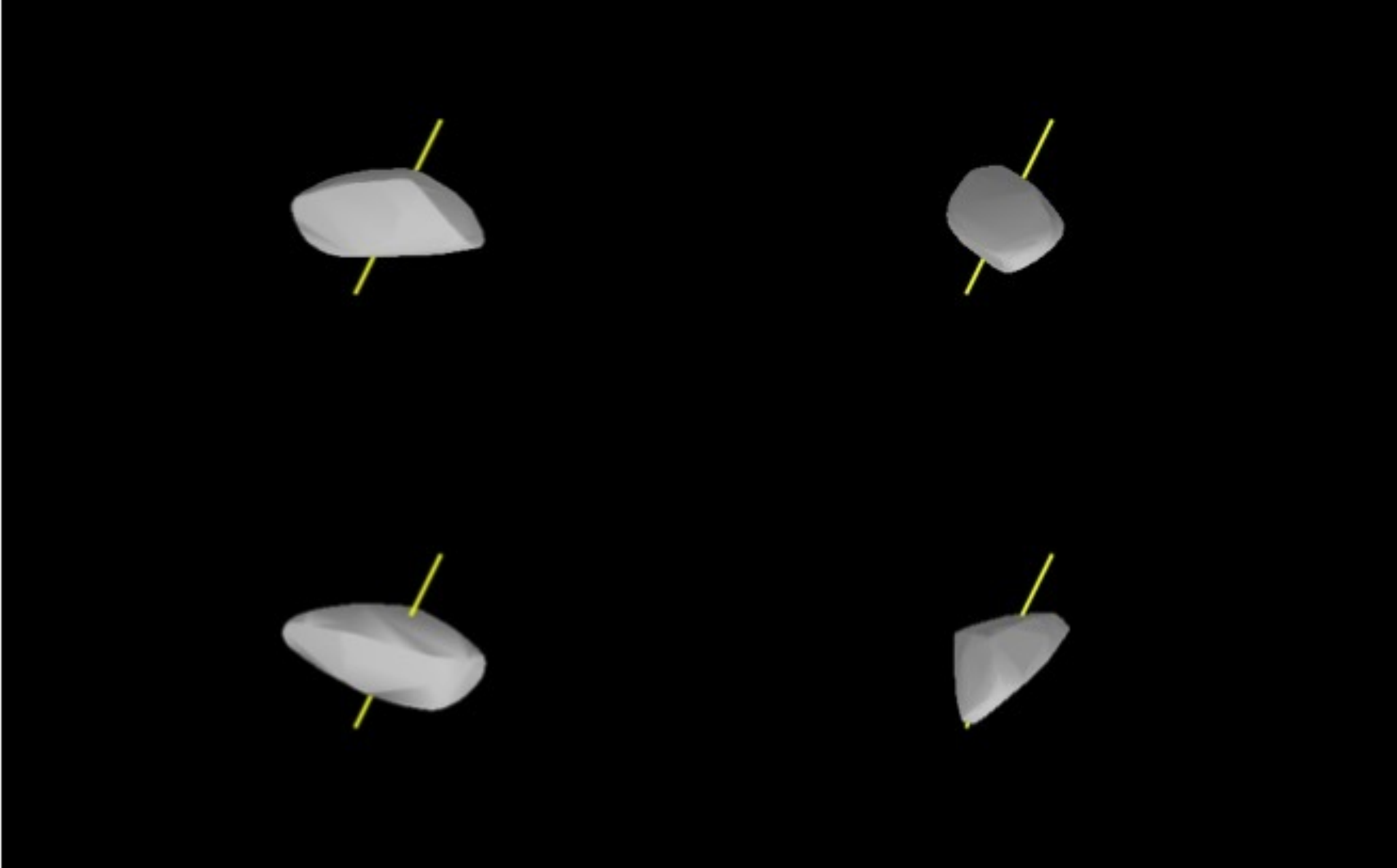}
   \caption{The output shape model of 206167 at four points about its rotation separated by $90^{\circ}$. The object is aligned at its obtained spin pole latitude $\beta=-67 \pm 5^{\circ}$.}
\label{fig:206167_model}
  \end{center}
 \end{figure}

Asteroid 206167 was observed by our program to have rotation period $P_{r} = 3.946 \pm 0.006$ and a maximum observed amplitude $A_{obs}= 1.07 \pm 0.05$. Additional observations of this object in July 2016 gave sufficient observations at different orbital geometries to allow a shape model to be determined.  The shape model obtained for this asteroid is shown in Figure~\ref{fig:206167_model} with best fit spin pole axes $\lambda= 57 \pm 5^{\circ}$, $\beta=-67 \pm 5^{\circ}$. 

\begin{figure}
  \begin{center}
\includegraphics[width=0.5\textwidth]{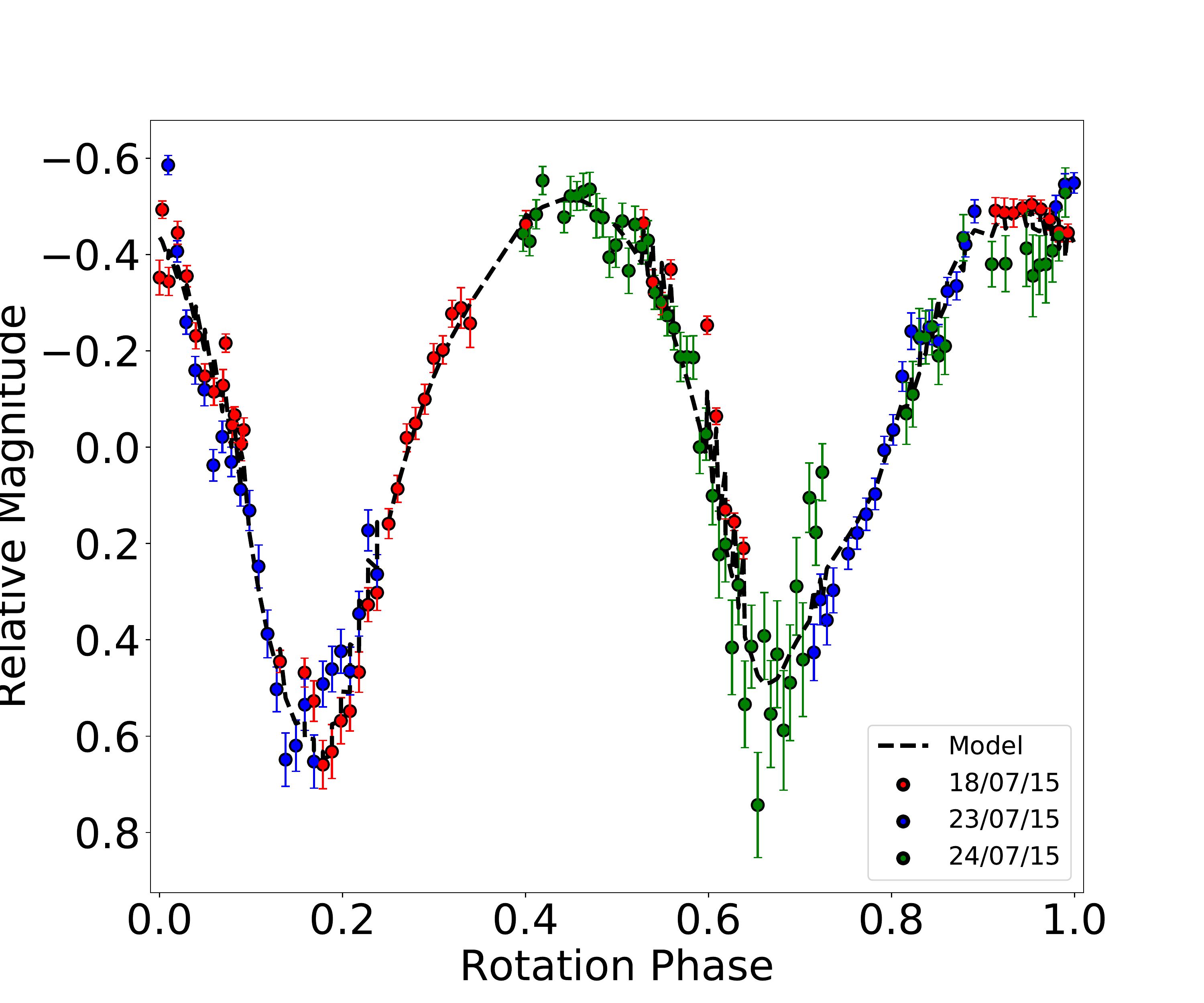}
   \caption{A light curve constructed from photometric data from observations of asteroid 206167 using the NTT in July 2015. The black dotted line is the light curve obtained from the shape model given in Figure~\ref{fig:206167_model} }
\label{fig:206167_fitlc}
  \end{center}
 \end{figure}

 The light curve produced by the shape model is plotted along with the July 2015 NTT observations in Figure~\ref{fig:206167_fitlc}. At the time of writing there are no shape or spin pole models for this object for comparison. 

\subsection{49257 (1998 TJ31)}
\label{subsec:inv_49257}

\begin{figure}
  \begin{center}
\includegraphics[width=0.5\textwidth]{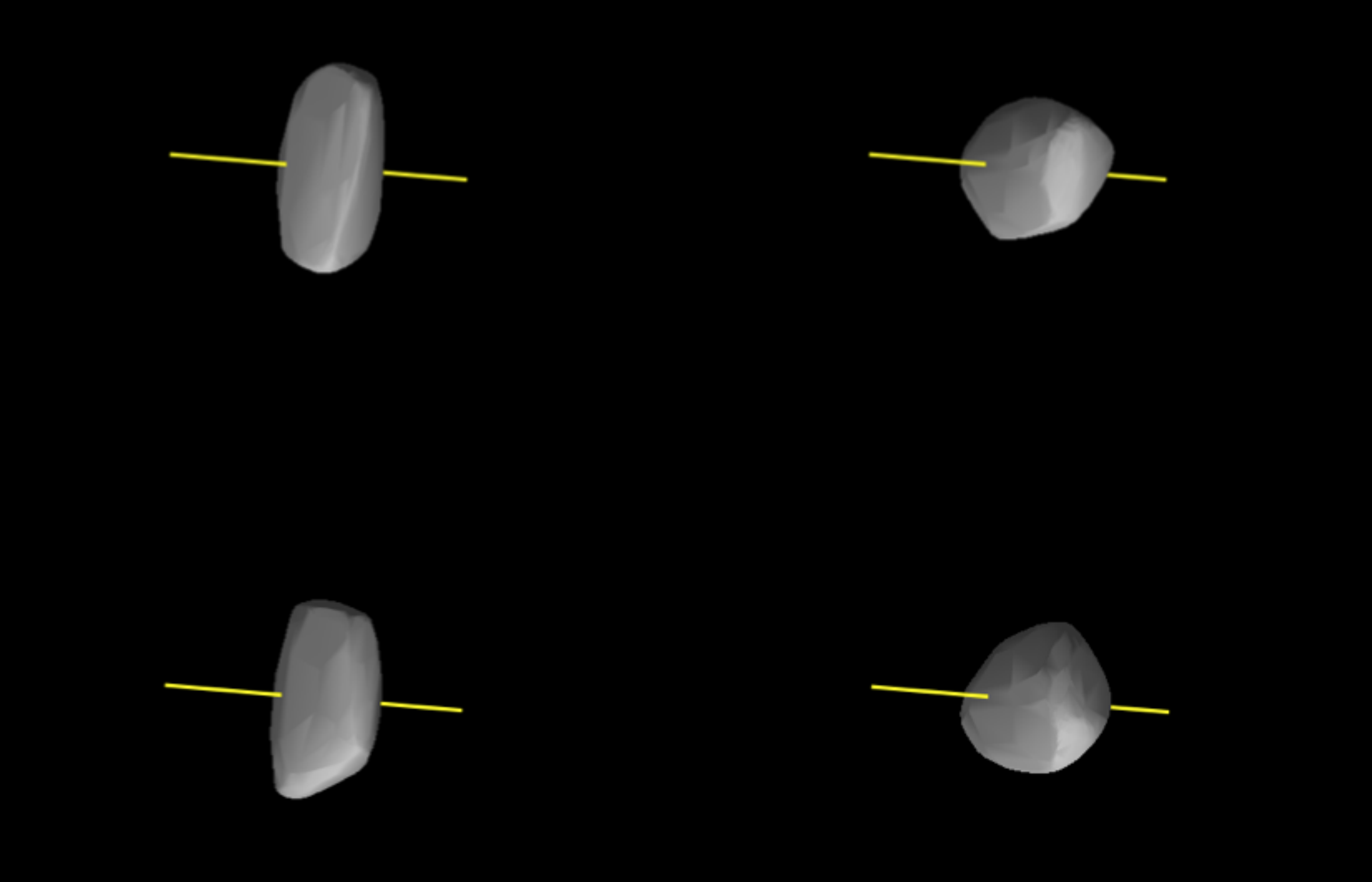}
   \caption{The output shape model of 49257 at four points about its rotation separated by $90^{\circ}$. The object is aligned at its obtained spin pole latitude $\beta=6^{\circ}$.}
\label{fig:49257_par}
  \end{center}
 \end{figure}

Asteroid 49257 was observed to have a light curve amplitude below the selection criterion decided for high amplitude objects. Despite this we serendipitously obtained further observations of this object at different orbital geometry which allowed us to obtain shape and spin data. The best fit shape result for this object is given in Figure~\ref{fig:49257_par} with best fit spin pole longitude and latitude $\beta= 112\pm 5^{\circ} , \theta=6\pm 5^{\circ}$.

The light curve produced by this model object is plotted along with the observed light curve from observations in July 2015 in Figure~\ref{fig:45864_lco1}.

\begin{figure}
  \begin{center}
\includegraphics[width=0.5\textwidth]{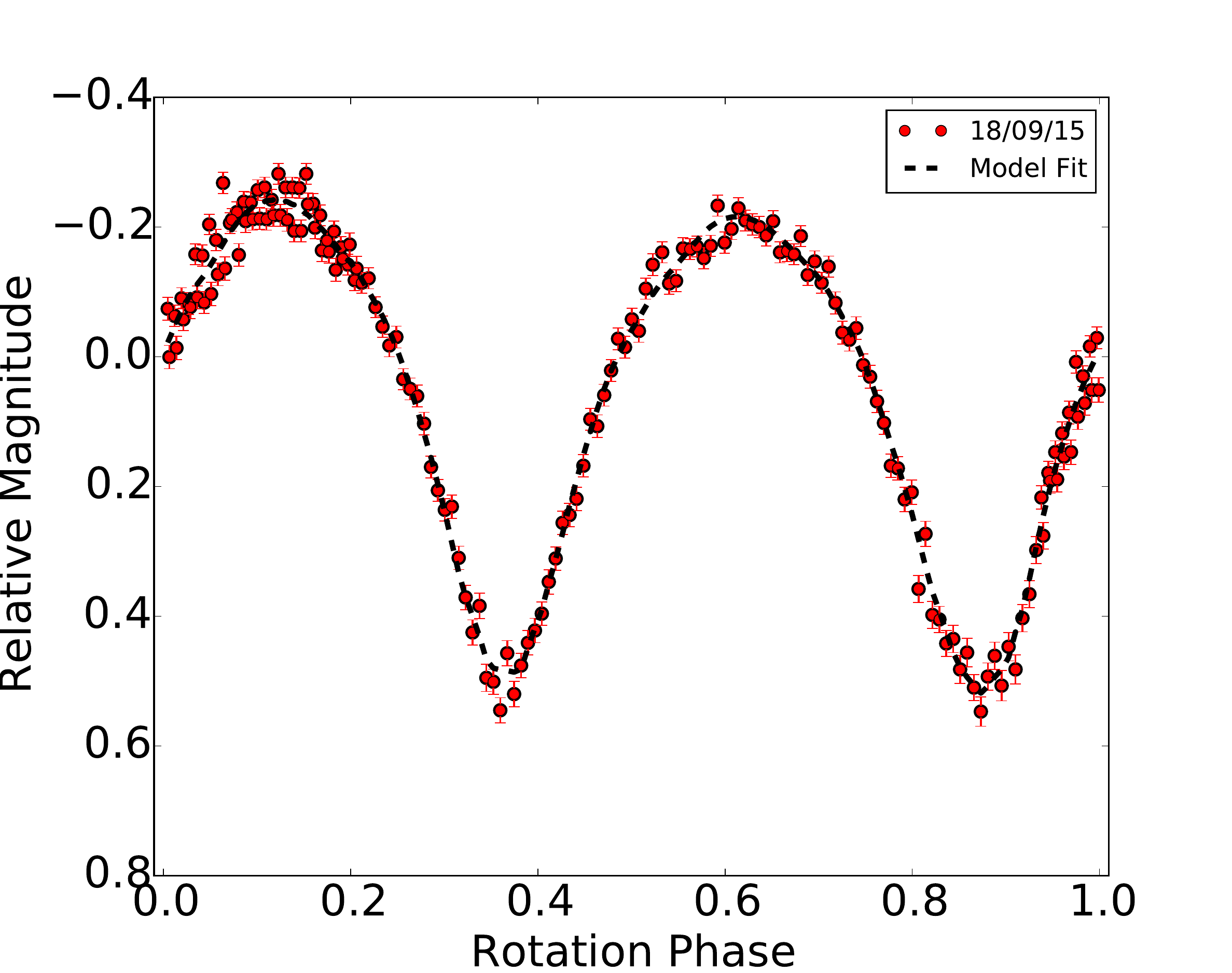}
   \caption{A light curve constructed from photometric data from observations of asteroid 49257 using the UH2.2m in September 2015. The black dotted line is the light curve obtained from the shape model given in Figure~\ref{fig:49257_par} }
\label{fig:49257_lc1}
  \end{center}
 \end{figure}

We see that the model shape fits the observed light curve fairly well in this case. It is also interesting to note that if this shape is accurate that if we observe the object 'edge-on' then we would expect to see a greater light curve amplitude. We hope to obtain further data for this object. At the time of writing there was no pre-existing shape or spin pole information for asteroid 49257 to use for comparison. The spin pole solution for this object suggests a high obliquity, \cite{lacerda2011} demonstrate that the light curve amplitude of a high obliquity object can vary greatly about its orbital geometry and this could explain why this object was not determined to be an 'extreme asteroid' despite the shape solution suggesting a higher amplitude than measured in our project.

\section{Conclusions}

Using the first 18 months of the PanSTARRS 1 survey we identified 33 candidate high-amplitude objects for follow-up observations and carried out observations of 22 asteroids. 4 of the observed objects have amplitude $A_{obs}\geq 1.0$ mag. Despite the fact that our selection criteria was initially defined for identification of asteroids with fast rotation periods, $P<2.2$ h we found none of these 'super-fast rotators' in our target list. Of the remaining observed objects for which light curve amplitudes $A_{obs} < 0.9$ mag were obtained, we are only able to definitively rule two out as high amplitude targets. The rotation periods we determine from the light curves of these asteroids were too long for the large variations seen in the PS1 survey data to be real without assuming an unrealistic amplitude. In future work we hope to obtain photometric colours and/or spectroscopy for these objects which will allow us to further constrain their density and strengths. Additionally knowing the taxonomic type of the asteroid will allow us to more accurately account for the phase-angle amplitude effect.


Asteroid 49257 was observed by our project to have an amplitude lower than our cut-off criteria. When we obtained enough further observational data for this objects to generate a shape model, however, we found that the best fit shape would be considered a high-amplitude asteroid.  At least one further set of observations at different orbital geometries is required to rule any of our targets out as high-amplitude asteroids. Assuming the shape distribution determined in \cite{mcneill2016} we find that as many as $0.6\%$ of main belt asteroids with $1<D<10$ km may be high amplitude objects with $A_{max} \geq 1.0$ mag. From this investigation we find 4 asteroids of 22 total observed to have amplitudes $A \geq 1.0$ mag. We can use this to estimate an expected lower limit on the abundance of high amplitude asteroids within survey data. Scaling the proportion of high amplitude asteroids observed to our master target list we estimate that from the full PS1 sample of $\sim 60,000$ objects we would expect $0.01\%$ to be high amplitude. This number is not particularly constraining as it only represents an estimate of the lower limit abundance of high amplitude asteroids.

We determine that 4 objects are likely to be single rubble pile objects with some cohesive strength allowing them to resist mass shedding even at their highly elongated shapes, dependent on their density. 3 further objects below the cut-off for 'high amplitude asteroids' had a combination of amplitude and period such that they may also require some density-dependent cohesive strength. None of these objects require any strength significantly greater than the determined values for known asteroids to date. It is also worth noting that increasing the angle of friction used in the calculations will reduce these values further (\citealt{sanchez2016}). Thus, although these objects display higher than average amplitude, due to the relatively small internal strengths required to resist rotational fission none are considered extraordinarily high amplitude.

For asteroids 45864, 206167 and 49257 in our program we had obtained enough data at different epochs and orbital geometries to allow us to model the shape and spin-pole orientation of these objects. 45864 was determined to have retrograde rotation with spin pole latitude and longitude $\lambda=218\pm 10^{\circ}, \beta=-82\pm 5^{\circ}$ and 206167 with $\lambda= 57 \pm 5^{\circ}$, $\beta=-67 \pm 5^{\circ}$. 49257 was determined to have a spin pole at $\lambda=112\pm 6^{\circ}, \beta=6\pm 5^{\circ}$. The high obliquity of 49257 could explain how we failed to identify this body as high amplitude from its light curve alone when its shape solution suggests otherwise as the light curve amplitude high obliquity objects is highly dependent on orbital geometry.

\section*{Acknowledgements}

We thank the referees for their feedback on this manuscript during the submission process which has improved the overall quality of the paper.

The Pan-STARRS1 Surveys (PS1) and the PS1 public science archive have been made possible through contributions by the Institute for Astronomy, the University of Hawaii, the Pan-STARRS Project Office, the Max-Planck Society and its participating institutes, the Max Planck Institute for Astronomy, Heidelberg and the Max Planck Institute for Extraterrestrial Physics, Garching, The Johns Hopkins University, Durham University, the University of Edinburgh, the Queen's University Belfast, the Harvard-Smithsonian Center for Astrophysics, the Las Cumbres Observatory Global Telescope Network Incorporated, the National Central University of Taiwan, the Space Telescope Science Institute, the National Aeronautics and Space Administration under Grant No. NNX08AR22G issued through the Planetary Science Division of the NASA Science Mission Directorate, the National Science Foundation Grant No. AST-1238877, the University of Maryland, Eotvos Lorand University (ELTE), the Los Alamos National Laboratory, and the Gordon and Betty Moore Foundation. We thank the PS1 Builders and PS1 operations staff for construction and operation of the PS1 system and access to the data products provided.

This work is based on observations made with the Isaac Newton Telescope and William Herschel Telescope operated on the island of La Palma by the Isaac Newton Group of Telescopes in the Spanish Observatorio del Roque de los Muchachos of the Instituto de Astrofisica de Canarias. Observations contributing to this project were made with the New Technology Telescope at the La Silla Observatory under programme ID 095.C-0336(A). 

The shape models given in this paper are displayed using MPO LCInvert. 

AM gratefully acknowledges support from DEL. AF acknowledges support from STFC research grant ST/L000709/1.

%

\vspace{5mm}
\facilities{Pan-STARRS 1; Isaac Newton Telescope, La Palma; New Technology Telescope, La Silla, Chile; University of Hawaii 2.2m Telescope}


\software{astropy \citep{astropy}}





\end{document}